\documentclass[
reprint,
amsmath,amssymb,
aps,prb
]{revtex4-2}

\usepackage{graphicx}
\usepackage{dcolumn}
\usepackage{bm}
\usepackage{color}
\usepackage{comment}

\usepackage{hyperref}
\hypersetup{
colorlinks=true,
citecolor=blue,
linkcolor=blue,
urlcolor=blue
}

\begin{document}

\preprint{APS/123-QED}

\title{
Universal description of massive point vortices and verification \\ 
methods of vortex inertia in superfluids
}

\author{Akihiro Kanjo$^{1}$}
 %\email{akihirokanjo@gmail.com}
\author{Hiromitsu Takeuchi$^{1,2}$}
 %\email{takeuchi@omu.ac.jp}
\affiliation{
$^1$ Department of Physics, Osaka Metropolitan University, 3-3-138 Sugimoto, Osaka 558-8585, Japan \\
$^2$ Department of Physics, Nambu Yoichiro Institute of Theoretical and Experimental Physics (NITEP), Osaka Metropolitan University, 3-3-138 Sugimoto, Osaka 558-8585, Japan
}

\date{\today}

%%%%%%%%%%%%%%%%%%%%%%%%%%%%%%%%%%%%%%%%%%%%%%%%%%%%%%%%%%%%%%%%%%%%%%%%%%%%%%%%%%%%%%%%%%%%%%%%%%%%%%%%%%%%%%%%

\begin{abstract}
Vortex mass, which is the inertia of a quantum vortex, has never been observed in superfluids and is a long-standing problem in low temperature physics.
The impact of the mass is considered negligible in typical experiments with superfluid $^4$He.
Recent developments of experimental techniques for manipulating quantum vortices in superfluid atomic gases have enabled us to test this problem more accurately.
By introducing the vortex mass time and length as universal scales to many-body problems of massive quantum vortices,
the theoretical description is formulated in the simplest manner and is universally applicable to different quantum fluids, including fermionic and multicomponent superfluids.
There are two branches, the cyclotron and massless branches, for the circular motion of a pair of like-sign vortices.
Finding a stable cyclotron branch for the motion of vortices is a clear evidence of vortex mass and superfluid $^3$He-B is the specific example of a system where this phenomena could be observed.
The impact of the mass on the massless branch is small but can be enhanced by taking the difference in the two-body dynamics of point vortices with different initial conditions.
Our results imply that the vortex mass is a direct cause of the splitting instability of a doubly quantized vortex at absolute zero and that
the vortex mass length
characterizes the final state after the instability.
It is also demonstrated that a pair of massive vortices with opposite circulations has a critical distance characterized by
the vortex mass length,
below which they are spontaneously annihilated without thermal fluctuations.
\end{abstract}

\maketitle

\section{\label{sec:introduction}Introduction}
A three-dimensional flow with vortices, which is generally complicated to treat analytically, is described in terms of vorticity rather than fluid velocity in hydrodynamics~\cite{batchelor1967introduction}. 
In classical hydrodynamics, a qualitative understanding of vortex dynamics is approximated by assuming that the continuous vorticity distribution is a collection of vortex filaments; that is, vortex tubes with infinitesimal cross sections~\cite{batchelor1967introduction}.
The vortex filament model has been successfully applied to the inviscid quantum liquids of superfluid $^4$He, where the circulation is quantized and vortices exist as stable topological defects with elementary circulation $\kappa$ defined by Planck constant over the mass of the condensation particles~\cite{donnelly1991quantized}.
Although this model is applicable to superfluid atomic gases, two-dimensional superfluid flows, rather than three-dimensional flows, have been widely studied because they enable us to observe the time evolution of vortices directly and control their motion in a sophisticated manner in highly oblate atomic clouds~\cite{freilich2010real,kwon2014relaxation,moon2015thermal,kwon2016observation}.
Two-dimensional vortex dynamics are described in the point vortex model (PVM)~\cite{saffman1995vortex} instead of the vortex filament model, where a collection of vortices is mathematically identified with a two-dimensional Coulomb gas~\cite{ragazzo1994motion,fischer1999motion}.

In superfluids, it is believed that a vortex has inertia, called vortex mass, which is typically neglected in classical hydrodynamics.
A quantum vortex forms the core, in which the superfluid order parameter vanishes.
Low-energy quasiparticles or elementary excitations are bound to the core of a vortex, and their localized population causes inertia in the vortex~\cite{kopnin1978frequency,kopnin1998dynamic,simula2018vortex}.
In multicomponent superfluids, the core of a quantum vortex in a superfluid component may be occupied by other components, the mass of which contributes to the vortex mass~\cite{richaud2020vortices,richaud2021dynamics,ruban2022direct,matteo2023massive,richaud2023massive,bellettini2024rotational,d2024stability,richaud2024suppression,richaud2024dynamical}.
If the vortex core can be modeled as an empty bubble,
then a virtual or induced mass is added to the above mass~\cite{baym1983hydrodynamics}.
The vortex mass is safely negligible in superfluid $^4$He because its core size is on the order of \AA \ and is much smaller than other relevant lengths.
In contrast, some theories suggest that the mass of a straight vortex is proportional to the logarithm of the system size over the core size~\cite{popov1973quantum,duan1992inertial,duan1994mass}, even though it can be logarithmically divergent in the typical setup of superfluid $^4$He.
Based on the similarity of its derivation to the electromagnetic mass, this type of logarithmic vortex mass is called the \textit{relativistic} mass~\cite{volovik2003universe}.
The nonlogarithmic mass introduced above is called the \textit{nonrelativistic} mass.
Despite these conflicting theories,
the vortex mass is a long-standing problem in low temperature physics and has not yet been qualitatively verified.

Recent improvements in experimental techniques motivated us to address this problem, particularly for superfluids with ultracold atoms.
First, the distance between the vortices is comparable to the vortex core size in typical experiments~\cite{moon2015thermal,kwon2021sound,hernandez2024connecting}.
In this situation, in contrast with the superfluid $^4$He, the vortex mass is expected to influence the many-body dynamics of quantum vortices.
Second, the vortex dynamics in a uniform superfluid can be observed in a two-dimensional box trap~\cite{kwon2021sound,hernandez2024connecting}. 
This contrasts with vortex dynamics in a conventional system trapped by a harmonic potential, where the density inhomogeneity makes it difficult to treat theoretically~\cite{shin2004dynamical,moon2015thermal}.
Finally, the initial condition of the point-vortex dynamics can be manipulated using the pinning potential~\cite{kwon2021sound,samson2016deterministic}.
The problem of vortex mass is now being discussed for all types of superfluids, from scalar to multicomponent superfluids of bosonic and fermionic condensates.
Unified theoretical descriptions and efficient observation methods for examining vortex mass are desired.

In this study, we investigate theoretically the dynamics of massive point vortices in uniform superfluids at zero temperature.
By introducing the characteristic time and length and using the resultant universal equation of motion, we focus on the simplest many-body problem: 
the dynamics of pair of vortices. 
In terms of the compressibility of fluids or the Mach number, we evaluate feasibility of the two circular orbit solutions of like-sign pairs for bosonic and fermionic superfluids.
It is shown that the vortex mass can be quantitatively measured by experimentally controlling the initial velocity along the circular orbit.
We also show that the vortex mass causes the splitting instability of a doubly quantized vortex at zero temperature.
Additionally, we investigate the dynamics of a vortex--antivortex, focusing on the oscillation in the relative direction.

This paper is organized as follows:
We formulate the equation of motion for massive point vortices with the vortex mass time and length in Sec.~\ref{sec:massive PVM}.
We also discuss the possible range of vortex mass, especially in ultracold atoms.
In Sec.~\ref{sec:vortex-vortex}, we present the main results of the dynamics of a vortex--vortex pair based on analytical and numerical calculations of trajectories depending on the initial conditions in bosonic and fermionic superfluids.
In Sec.~\ref{sec:vortex-antivortex}, we investigate the dynamics of a vortex--antivortex pair in the same manner as that of a vortex--vortex pair.
The summary and discussion are presented in Sec.~\ref{sec:conclusion}.

\section{\label{sec:massive PVM}Massive point vortex model}

\subsection{\label{sec:formulation}Formulation}
We consider the two-dimensional dynamics of $N$ vortices without a background flow.
In the conventional PVM~\cite{saffman1995vortex}, the $i$th vortex at the position $\bm{r}_i$ moves along with the sum of the flows induced by all the other vortices:
\begin{equation}
    \frac{d\bm{r}_i(t)}{dt} 
    =\sum^N_{j\neq i} \bm{v}_{ij}
    \label{eq:massless PVM}
\end{equation}
with induced velocity
\begin{equation*}
    \bm{v}_{ij}
    =\frac{\Gamma_j}{2\pi}\frac{\bm{e}_z\times(\bm{r}_i-\bm{r}_j)}{|\bm{r}_i-\bm{r}_j|^2},
\end{equation*}
where $\Gamma_j$ is the circulation of the $j$th vortex and $\bm{e}_z$ is the unit vector along the $z$ axis.
The circulation of the most stable vortex is $\kappa$, which is the circulation quantum, and we set $\Gamma_j=\pm\kappa$.
Suppose an isolated pair of vortices with a distance $r=|\bm{r}_2-\bm{r}_1|$ and the same circulation ($\Gamma_1=\Gamma_2=\kappa$).
These vortices will rotate about the vortex centroid $\bm{r}_G=(\bm{r}_1+\bm{r}_2)/2$ with an angular velocity
\begin{equation}
    \Omega_0(r)=\frac{\kappa}{\pi r^2}.
    \label{eq:omega0}
\end{equation}
If the circulations are in opposite directions ($\Gamma_1=-\Gamma_2=\kappa$), the pair does not rotate but moves through the fluid with a centroid velocity 
\begin{equation}
    V_0(r)=\frac{\kappa}{2\pi r}.
    \label{eq:V0}
\end{equation}

The massive PVM corresponds to a system of charged cylinders with logarithmic potential in a uniform magnetic field~\cite{ragazzo1994motion,fischer1999motion}.
In this analogy to the conventional PVM, the charged cylinders are massless, and accordingly, the vortices have no inertia.
We then write the equation of motion for the massive point vortices as
\begin{equation}
    m_i \frac{d^2\bm{r}_i}{dt^2}
    =\rho\Gamma_i\bm{e}_z\times
     \left(\frac{d\bm{r}_i}{dt}-\sum^N_{j\neq i}\bm{v}_{ij}\right),
     \label{eq:massive PVM}
\end{equation}
where $m_i$ is the mass of the $i$th vortex per unit length and $\rho$ is the fluid density. 
The right side of Eq.~\eqref{eq:massive PVM} represents the Magnus force in hydrodynamics that acts when the velocity $d\bm{r}_i/dt$ of the $i$th vortex is not equal to the sum of the induced velocities.
Equation~\eqref{eq:massive PVM} without vortex mass ($m_i=0$) is reduced to Eq.~\eqref{eq:massless PVM}.

Here, we discuss the conditions under which the massive PVM is applicable from the perspective of fluid compressibility.
In hydrodynamics~\cite{batchelor1967introduction}, compressibility is typically neglected if the Mach number $M$, which is the ratio of the relative motion speed $u$ with respect to the local flow velocity and speed of sound $c_s$, fulfills the condition
\begin{equation}
    M\equiv\frac{u}{c_s}\lesssim0.3.
    \label{eq:Mach number}
\end{equation}
In our case, $u$ corresponds to the vortex velocity relative to the sum of induced velocities;
\begin{equation}
    u=\left|\frac{d\bm{r}_i}{dt}-\sum^N_{j\neq i}\bm{v}_{ij}\right|.
    \label{eq:relative motion speed}
\end{equation}
In this study, we simply combine condition~\eqref{eq:Mach number} with Eq.~\eqref{eq:relative motion speed} to evaluate the applicability of massive PVM in superfluid systems.

To gain insight into the massive PVM, we clarify the correspondence in the analogy.
By substituting $q_i\to\sqrt{\varepsilon_0\rho}\Gamma_i$ and $B\to\sqrt{\rho/\varepsilon_0}$, we obtain the equation of motion for a two-dimensional Coulomb gas under a uniform magnetic field $-B\bm{e}_z$, as follows:
\begin{equation}
    m_i \frac{d^2\bm{r}_i}{dt^2}
    =q_i \frac{d\bm{r}_i}{dt}\times (-B)\bm{e}_z
    +\sum^N_{j\neq i} \frac{q_iq_j}{2\pi\varepsilon_0} \frac{\bm{r}_i-\bm{r}_j}{|\bm{r}_i-\bm{r}_j|^2}.
    \label{eq:EOM for charges}
\end{equation}
Here, $m_i$ and $q_i$ are the mass and charge of the $i$th particle, respectively, and $\varepsilon_0$ represents the dielectric constant in vacuum.
The first and second terms on the right side of Eq.~\eqref{eq:EOM for charges} represent the Lorentz force and Coulomb interactions, respectively.
In the case of point vortices, the Magnus force consists of counterparts of the Lorentz force and Coulomb interactions.
The former contains the self-velocity $d\bm{r}_i/dt$ and is named the self-Magnus term.
The latter represents the interaction between vortices; therefore, we refer to this as the mutual Magnus term.

To systematically describe our system, the universal time and length scales should be introduced.
For simplicity, we assume that each quantum vortex has the same vortex mass $m_i=m_v$.
According to the above analogy, the one-body vortex dynamics obey cyclotron motion owing to the self-Magnus term: 
thus, the universal timescale is the inverse of the cyclotron frequency $|q_i|B/m_i\to\rho\kappa/m_v$ and is given by
\begin{equation}
    \tau=\frac{m_v}{\rho\kappa}.
    \label{eq:vortex mass time}
\end{equation}
The universal length scale is associated with the many-body effect or the mutual Magnus term
although no counterpart has been introduced in the literature in contrast with $\tau$.
By using $m_v$ and $\rho$, it is introduced by
\begin{equation}
    \sigma=\sqrt{\frac{m_v}{\pi\rho}},
    \label{eq:vortex mass length}
\end{equation}
where $\pi$ is included for convenience of discussion.
In this paper, we name $\tau$ and $\sigma$ as the vortex mass time and the vortex mass length, respectively.

Eventually, we can perform a universal analysis of systems of massive quantum vortices by rendering the equation of motion dimensionless in terms of $\tau$ and $\sigma$:
this is because the resulting dimensionless equation contains no explicit parameters.
For our two-vortex problem ($N=2$), by rescaling Eq.~\eqref{eq:EOM for charges} with
\begin{equation*}
    \tilde{t}=\frac{t}{\tau},\quad
    \tilde{\bm{r}}_i=\frac{\bm{r}_i}{\sigma},
\end{equation*}
we obtain the following coefficientless equation: 
\begin{equation}
    \frac{d^2\tilde{\bm{r}}_i}{d\tilde{t}^2}
    =\pm\bm{e}_z\times\frac{d\tilde{\bm{r}}_i}{d\tilde{t}}
    \pm\frac{1}{2}\frac{\tilde{\bm{r}}_i-\tilde{\bm{r}}_j}{|\tilde{\bm{r}}_i-\tilde{\bm{r}}_j|^2}
    \quad (i,j=1,2).
    \label{eq:dimensionless EOM}
\end{equation}
Similar to Newtonian mechanics, the vortex motion is determined by the initial conditions $\tilde{\bm{r}}_1(0)$, $\tilde{\bm{r}}_2(0)$, $d\tilde{\bm{r}}_1(0)/d\tilde{t}$, and $d\tilde{\bm{r}}_2(0)/d\tilde{t}$.
Our problem is substantially simpler than those in previous studies~\cite{ragazzo1994motion,richaud2021dynamics,khalifa2024vortex}.
Regardless of this simplification, vortex dynamics are still complicated in the presence of the self-Magnus term compared with conventional mass point systems in Newtonian mechanics.
The positive (negative) sign of the first term on the right side of Eq.~\eqref{eq:dimensionless EOM} denotes that the circulation of the $i$th vortex is $+\kappa$ ($-\kappa$). 
The positive (negative) sign of the second term corresponds to the problem of a vortex--(anti)vortex pair, as discussed in Sec.~\ref{sec:vortex-vortex} (Sec.~\ref{sec:vortex-antivortex}).

\subsection{\label{sec:vortex mass}Possible range of vortex mass}
A simple model for a nonrelativistic mass considers a quantum vortex a cylinder of radius $r_{\mathrm{core}}$.
In this case, the vortex mass equals the mass of the cylinder and virtual hydrodynamic mass.
The latter is predicted by hydrodynamics and is given by $\pi\rho r_{\mathrm{core}}^2$ with fluid density $\rho$~\cite{baym1983hydrodynamics}. 
The former, namely the vortex-internal mass, is zero if the vortex core is empty.
Quasiparticles bound at the core can make this contribution nonzero in fermionic superfluids~\cite{kopnin1978frequency,kopnin1998dynamic}, which is of the same order as $\pi\rho\xi^2$ with healing length $\xi$~\cite{volovik1998vortex,volovik2003universe}.
Because the core size $r_{\mathrm{core}}$ of a quantum vortex is of the order of $\xi$, the internal vortex mass is comparable to the virtual hydrodynamic mass.
In bosonic superfluids, it has also been suggested that the vortex internal mass is induced by the Kelvin modes or the kelvon quasiparticles bounded at the core~\cite{simula2018vortex}.
The internal vortex mass can be larger than the virtual hydrodynamic mass in immiscible binary superfluids~\cite{richaud2020vortices,richaud2021dynamics,ruban2022direct,matteo2023massive,richaud2023massive,bellettini2024rotational,d2024stability,richaud2024suppression,richaud2024dynamical}.
The core of the vortex in the first superfluid component is filled by the second component by forming a coreless vortex.
The internal mass of the vortex is proportional to the population of the second component and is estimated as $\pi\rho_{\mathrm{core}}r_{\mathrm{core}}^2$, where $\rho_{\mathrm{core}}$ denotes the mass density of the second component.

The relativistic mass is related to the hydrodynamic energy of the quantum vortex.
Analogous to electromagnetics, the counterpart to the rest mass is a candidate for vortex mass~\cite{popov1973quantum,volovik2003universe}.
This is defined as $m_{\mathrm{rel}}=E_{\mathrm{vor}}/c_s^2\sim(\rho\kappa^2/4\pi c^2_s)\ln R_{\mathrm{sys}}/\xi$ with system size $R_{\mathrm{sys}}$.
Here, $E_{\mathrm{vor}}$ is the energy per unit of length and is associated with the superfluid velocity field of a static vortex.
When $R_{\mathrm{sys}}$ is sufficiently large, the relativistic mass diverges logarithmically.
Furthermore, the mass can change according to vortex motion because hydrodynamic energy depends on the position and interactions with other vortices.

To estimate the possible range of the vortex mass, we ignore the variation in the relativistic mass caused by vortex motion.
The nonrelativistic mass is typically of the same order as $\pi\rho\xi^2$.
The relativistic mass $m_{\mathrm{rel}}$ is of the same order as $\pi\rho\xi^2$ multiplied by the factor $\ln R_{\mathrm{sys}}/\xi$. 
For large condensates~\cite{moon2015thermal}, for example, 
we obtain that $\ln R_{\mathrm{sys}}/\xi\approx\ln 76/0.3\approx5.53$.
In the following analysis, we reasonably choose the range of the vortex mass as
\begin{equation}
    0<\frac{m_v}{\pi\rho\xi^2}=\left(\frac{\sigma}{\xi}\right)^2\lesssim10.
    \label{eq:range of vortex mass}
\end{equation}
Furthermore, it is imorotant that the impact of vortex mass on the many-body vortex dynamics can be evaluated universally for different superfluid systems in terms of the ratio of the vortex mass length $\sigma$ to the healing length $\xi$.
To clarify the effect of mass, we first consider a uniform system without a boundary.

\section{\label{sec:vortex-vortex}A vortex--vortex pair}
In this section, we consider the dynamics of a pair of quantum vortices with the same circulation, $\Gamma_1=\Gamma_2=\kappa$.
The problem is formulated in Sec.~\ref{sec:effective radial potential}, and the significant results are presented in the following sections.
In Sec.~\ref{sec:circular orbits}, the two solutions of the relative circular orbits are evaluated in terms of the Mach number.
In Sec.~\ref{sec:angle shift}, we propose an efficient method for experimentally detecting vortex mass.
Our results relate to the splitting instability of a doubly quantized vortex in Sec.~\ref{sec:instability}.

\subsection{\label{sec:effective radial potential}Effective radial potential}
For the problem of a vortex--vortex pair, the equation of motion is given by Eq.~\eqref{eq:dimensionless EOM} with the plus sign of the first and second terms.
To clarify its correspondence with Newtonian mechanics, we begin with an equation without universal rescaling.
We introduce the center of mass coordinate $\bm{r}_G=(\bm{r}_1+\bm{r}_2)/2$ and the relative coordinate $\bm{r}_R=\bm{r}_2-\bm{r}_1$.
The two-body vortex problem can then be reduced to two independent one-body vortex problems: 
\begin{align}
    M_v\frac{d^2\bm{r}_G}{dt^2} 
    &=2\rho\kappa\bm{e}_z\times\frac{d\bm{r}_G}{dt},
    \label{eq:EOM for vortex-vortex,G}\\
    \mu_v\frac{d^2\bm{r}_R}{dt^2} 
    &=\frac{\rho\kappa}{2}\bm{e}_z\times\frac{d\bm{r}_R}{dt}+\frac{\rho\kappa^2}{2\pi}\frac{\bm{r}_R}{|\bm{r}_R|^2},
    \label{eq:EOM for vortex-vortex,R}
\end{align}
where $M_v=2m_v$ and $\mu_v=m_v/2$ are the total and reduced mass, respectively.
The general solution to Eq.~\eqref{eq:EOM for vortex-vortex,G} is 
\begin{align*}
    x_G(t)
    &=\tau v_G(0)\cos\left(\frac{t}{\tau}+\phi\right)+(x_G(0)-\tau v_G(0)\cos\phi), \\ 
    y_G(t)
    &=\tau v_G(0)\sin\left(\frac{t}{\tau}+\phi\right)+(y_G(0)-\tau v_G(0)\sin\phi)
\end{align*}
with Eq.~\eqref{eq:vortex mass time} and the initial conditions $\bm{r}_G(0)=(x_G(0), y_G(0))$ and $\dot{\bm{r}}_G(0)=(-v_G(0)\sin\phi, v_G(0)\cos\phi)$.
Thus, the center of mass motion is
a uniform circular motion of frequency $\tau^{-1}$
and radius $\tau v_G(0)$,
corresponding to the cyclotron motion of a particle with charge $\sqrt{\varepsilon_0\rho}\kappa$ in a uniform magnetic field $-\sqrt{\rho/\varepsilon_0}\bm{e}_z$. 
For simplicity, we set $v_G(0)=0$ and $x_G(0)=y_G(0)=0$ and ignore the center of mass motion, $\bm{r}_G(t)=0$, below.

\begin{figure}[htbp]
\centering
\includegraphics[width=8.45cm]{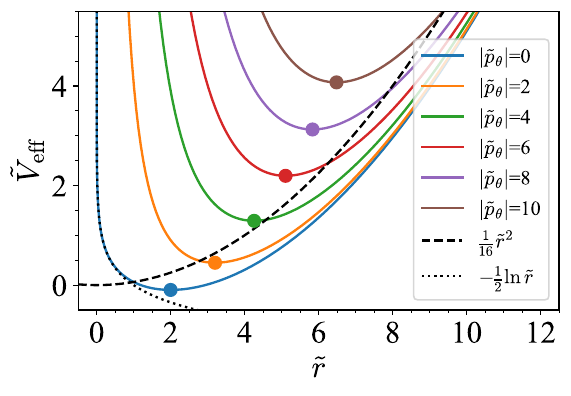}
\caption{ 
Plots of the effective radial potential $\tilde{V}_{\mathrm{eff}}$ (solid) for several values of $|\tilde{p}_\theta|$.
Dashed and dotted guidelines show the harmonic potential by the self-Magnus term and the logarithmic interaction potential by the mutual Magnus term, respectively.
Markers represent the local minimum point for each $|\tilde{p}_\theta|$.
}
\label{fig:potential for vortex-vortex}
\end{figure}

Equation~\eqref{eq:EOM for vortex-vortex,R} describes the relative motion.
The first term on the right represents the self-Magnus term.
The second term represents the mutual Magnus term, which is nonlinear, and thus complicates the dynamics.
To obtain a deeper physical insight, we introduced the Lagrangian of relative motion
\begin{equation}
    L_R(\bm{r}_R,\dot{\bm{r}}_R)
    =\frac{1}{2}\mu_v\dot{\bm{r}}_R^2
    +\frac{\rho\kappa}{4}\bm{e}_z\cdot(\dot{\bm{r}}_R\times\bm{r}_R)
    +\frac{\rho\kappa^2}{2\pi}\ln\frac{|\bm{r}_R|}{\sigma},
    \label{eq:L_R}
\end{equation}
where a constant is added and included in the last term.
Using the polar coordinates $\bm{r}_R=(r,\theta)$, the Lagrangian transforms into 
\begin{equation*}
    L_R(r,\dot{r},\theta,\dot{\theta})
    =\frac{1}{2}\mu_v(\dot{r}^2+r^2\dot{\theta}^2)
    -\frac{\rho\kappa}{4}r^2\dot{\theta}
    +\frac{\rho\kappa^2}{2\pi}\ln\frac{r}{\sigma}.
\end{equation*}
Here, the canonical momenta are
\begin{align}
    p_r
    &=\frac{\partial L_R}{\partial \dot{r}}
     =\mu_v\dot{r},
     \label{eq:radial momentum} \\
    p_\theta 
    &=\frac{\partial L_R}{\partial \dot{\theta}}
     =\mu_vr^2\dot{\theta}-\frac{\rho\kappa}{4}r^2.
     \label{eq:azimuth momentum}
\end{align}
The first term on the right side of Eq.~\eqref{eq:azimuth momentum} denotes the mechanical angular momentum. 
The second term is derived from the self-Magnus term.
Hence, $p_\theta$ is called a pseudomomentum.
The Hamiltonian is defined by the Legendre transformation $H_R=\dot{r}p_r+\dot{\theta}p_\theta-L_R$ and we obtain
\begin{equation}
    H_R(r,\theta,p_r,p_\theta)
    =\frac{p_r^2}{2\mu_v}+V_{\mathrm{eff}}(r,p_\theta)
    \label{eq:Hamiltonian for vortex-vortex}
\end{equation}
with effective radial potential
\begin{equation}
    V_{\mathrm{eff}}(r,p_\theta)
    =\frac{p_\theta^2}{2\mu_v r^2}
    +\frac{\rho\kappa}{4\mu_v}p_\theta
    +\frac{\rho^2\kappa^2}{32\mu_v}r^2
    -\frac{\rho\kappa^2}{2\pi}\ln\frac{r}{\sigma}.
    \label{eq:effective radial potential}
\end{equation}
Hamilton's equations are as follows:
\begin{align*}
    \dot{r}
    &=\frac{\partial H_R}{\partial p_r}
     =\frac{p_r}{\mu_v}, \\
    \dot{\theta}
    &=\frac{\partial H_R}{\partial p_\theta}
     =\frac{\partial V_{\mathrm{eff}}}{\partial p_\theta}, \\
    \dot{p}_r
    &=-\frac{\partial H_R}{\partial r}
     =-\frac{\partial V_{\mathrm{eff}}}{\partial r}, \\
    \dot{p}_\theta
    &=-\frac{\partial H_R}{\partial \theta}
     =0. 
\end{align*}
The last equation shows that $p_\theta$ is conserved and can be treated as a parameter.
The first term on the right side of Eq.~\eqref{eq:effective radial potential} denotes the centrifugal potential.
The second term is a constant.
The third term represents the harmonic potential arising from the self-Magnus term.
The last term represents the logarithmic interaction potential of the mutual Magnus terms.
Therefore, the relative motion is finally reduced to one-dimensional motion in the effective radial potential $V_{\mathrm{eff}}$.

To reveal the structure of $V_{\mathrm{eff}}$, we investigated the position of the extremum, $d V_{\mathrm{eff}}/d r=0$ at $r=r_*$, where
\begin{equation}
    r_*(p_\theta)
    \equiv\sigma\sqrt{2+\sqrt{4+\left(4\frac{p_\theta}{m_v\sigma^2/\tau}\right)^2}}
    \geq2\sigma
    \label{eq:local minimum}
\end{equation}
with $\mu_v=m_v/2$.
In Eq.~\eqref{eq:local minimum}, the equality holds for $p_\theta=0$, and $r_*$ has a minimum $2\sigma$~\cite{ruban2022direct}. 
As $d^2 V_{\mathrm{eff}}(r_*)/d r^2>0$ holds for any $p_\theta$, $V_{\mathrm{eff}}$ exhibits a local minimum at $r=r_*$.
Using $m_v$, $\tau$, and $\sigma$, $V_{\mathrm{eff}}$ is rescaled to the simplest form
\begin{equation*}
    \tilde{V}_{\mathrm{eff}}
    \equiv\frac{V_{\mathrm{eff}}-\frac{\rho\kappa}{4\mu_v}p_\theta}{m_v(\sigma/\tau)^2}
    =\frac{\tilde{p}_\theta^2}{\tilde{r}^2}
    +\frac{1}{16}\tilde{r}^2
    -\frac{1}{2}\ln\tilde{r}, 
\end{equation*}
where the constant term is subtracted for convenience, with
\begin{align*}
    \tilde{r}
    &\equiv\frac{r}{\sigma}, \\
    \tilde{\dot{\theta}}
    &\equiv\tau\dot{\theta}, \\
    \tilde{p}_\theta 
    &\equiv\frac{p_\theta}{m_v\sigma^2/\tau}
    =\frac{\tilde{r}^2\tilde{\dot{\theta}}}{2}-\frac{\tilde{r}^2}{4}.
\end{align*}
For reference, typical plots of $\tilde{V}_{\mathrm{eff}}$ and the local minimum points are shown for several values of $|\tilde{p}_\theta|$ in Fig.~\ref{fig:potential for vortex-vortex}.

Finally, we review the finite-size effect on the dynamics of massive point vortices (see Ref.~\cite{richaud2020vortices}).
Supposing that the condensates are trapped in a circular box potential, we consider the dynamics of a vortex--vortex pair in superfluids with a round boundary of radius $R_{\mathrm{sys}}$.
By assuming an initial condition with the twofold rotational symmetry of the system with respect to the origin (see Appendix~\ref{app:finite size effect}), we obtain the modified Hamiltonian of the relative motion 
\begin{equation}
    H'_R=H_R+\frac{\rho\kappa^2}{2\pi}\ln\left[1-\left(\frac{r}{2R_{\mathrm{sys}}}\right)^4\right],
    \label{eq:H'_R}
\end{equation}
where $H_R$ denotes the Hamiltonian without boundaries, given by Eq.~\eqref{eq:Hamiltonian for vortex-vortex}.
The logarithmic term in the second term on the right side is added to $V_{\mathrm{eff}}$ in $H_R$ and the finite-size effect is summarized by the factor $(r/2R_{\mathrm{sys}})^4$.
Therefore, if the ratio of the intervortex distance to the radius of the condensate is less than 0.31, the contribution of the finite-size effect becomes less than 1\% and is therefore negligible.

\subsection{\label{sec:circular orbits}Circular orbit solutions}
\begin{figure}[tbp]
\centering
\includegraphics[width=8.45cm]{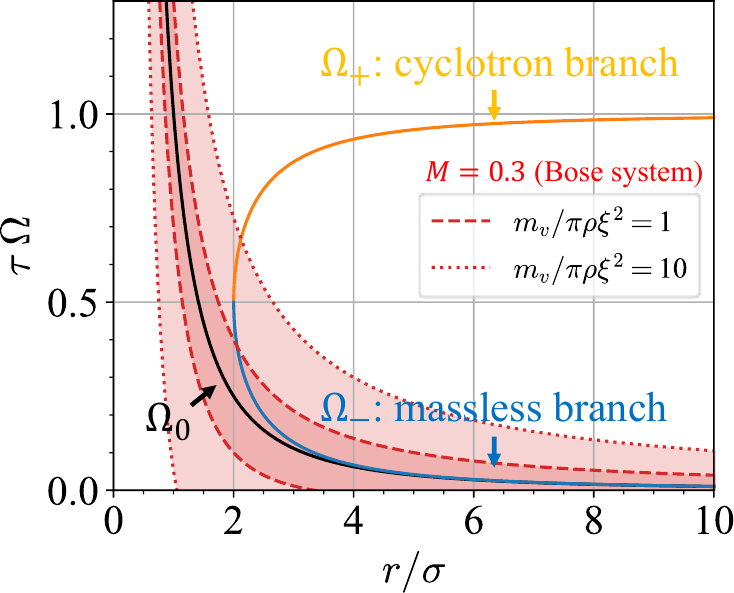}
\caption{
Plots of the angular frequencies $\Omega_{\pm,0}(r)$ (solid) rescaled with $\tau$ and $\sigma$.
The orange, blue, and black curves correspond to $\Omega_+$, $\Omega_-$, and $\Omega_0=\frac{1}{\tau(r/\sigma)^2}$, respectively [see Eqs.~\eqref{eq:omega} and \eqref{eq:omega0}].
The red regions represent the applicable condition of the massive PVM given by Eq.~\eqref{eq:Mach number for omega} with the boundary $M=0.3$ for $m_v/\pi\rho\xi^2=1$ (dashed) and $m_v/\pi\rho\xi^2=10$ (dotted) in bosonic superfluids, using $\tilde{c}_s^B=\frac{1}{2}(m_v/\pi\rho\xi^2)^{1/2}$.
In fermionic superfluids, the lower boundary is in the vicinity of $r/\sigma=0$, and the upper boundary is typically far above the plot area.
}
\label{fig:angular frequency}
\end{figure}

Due to the conservation of $p_\theta$, we find that $\dot{\theta}(t)$ is a constant, according to Eq.~\eqref{eq:azimuth momentum} when $r(t)$ is constant.
As $V_{\mathrm{eff}}$ has the local minimum at $r=r_*$ for any $p_\theta$, it is clear that there is the circular orbit solution with $r(t)=r_*$ and $\dot{r}(t)=0$.
Substituting Eq.~\eqref{eq:azimuth momentum} into Eq.~\eqref{eq:local minimum} and solving this problem with respect to $\dot{\theta}=\Omega$, we have 
\begin{equation}
    \Omega
    =\Omega_\pm(r_*)
    \equiv\frac{1}{2\tau}\left[1\pm\sqrt{1-\frac{4}{(r_*/\sigma)^2}}\right].
    \label{eq:omega}
\end{equation}
We call these two branches with the plus and minus signs the cyclotron and massless branches, respectively.
From Eq.~\eqref{eq:local minimum}, $\Omega_\pm$ is always real 
and it is straightforward to prove that these circular orbit solutions are dynamically stable
by investigating the linear stability of them.
The graphs of $\Omega_\pm(r)$ and $\Omega_0(r)$ are shown in Fig.~\ref{fig:angular frequency}.

Although these solutions have already been demonstrated in previous studies~\cite{richaud2020vortices,richaud2021dynamics},
we examine them in more detail to reveal the physical meaning of the two branches. 
The massless branch $\Omega_-$ is asymptotic to frequency $\Omega_0$ of the massless PVM for $r_*/\sigma\gg1$.
The cyclotron branch $\Omega_+$ approaches the cyclotron frequency $\tau^{-1}$ for $r_*/\sigma\gg1$.
This behavior of $\Omega_+$ is attributed to the fact that the mutual Magnus term owing to the intervortex interaction is less important for a large distance $r_*$ and the motion reduces to a one-body cyclotron motion.
The two branches have the same frequency $\Omega_\pm=2\Omega_0$ at $r_*=2\sigma$.
There exists no solution for a circular orbit with $r_*<2\sigma$.
These results are discussed in detail in Sec.~\ref{sec:instability}, which is related to the splitting instability of a doubly quantized vortex.

The stability of the two branches can be evaluated by computing the Mach number.
If we fix an object in a flow with a higher Mach number, the fluid compressibility is important, and predictions of incompressible hydrodynamics fail.
For a circular orbit with frequency $\Omega$ and radius $r$, the relative vortex speed is $u=\left|\frac{r}{2}\Omega-\frac{r}{2}\Omega_0\right|$, using Eq.~\eqref{eq:relative motion speed}.
According to Eq.~\eqref{eq:Mach number}, a circular orbit solution with frequency $\Omega(r)$ must satisfy the following conditions:
\begin{equation}
    M=\frac{r}{2c_s}|\Omega(r)-\Omega_0(r)|\lesssim0.3
    \label{eq:Mach number for omega}
\end{equation}
with $c_s=\tilde{c}_s\sigma/\tau$.
Hence, the solutions for the cyclotron and massless branches are valid when $\Omega=\Omega_{\pm}(r)$ satisfies this condition.
In other words, even if we could prepare the initial condition of the circular orbit with $\Omega_\pm$ outside the condition~\eqref{eq:Mach number for omega}, the vortices experience an extra drag force owing to the emission of sound (or quasiparticles), and the velocity of each vortex decays to a smaller value than the initial one.

The condition~\eqref{eq:Mach number for omega} was quantitatively evaluated for bosonic and fermionic superfluids.
The speed of the first sound in bosonic superfluids is $c_s\sim\kappa/2\pi\xi$ and is rescaled as  
\begin{equation*}
    \tilde{c}_s^B
    =\frac{c_s^B}{\sigma/\tau}
    \sim\sqrt{\frac{m_v}{\pi\rho\xi^2}}.
\end{equation*}
This holds for both superfluid $^4$He and atomic Bose--Einstein condensates (BECs)
~\cite{donnelly1991quantized,pethick2008bose}.
As shown in Fig.~\ref{fig:angular frequency}, $\Omega=\Omega_-$ ($\Omega=\Omega_+$) is typically included (excluded) in the range given by Eq.~\eqref{eq:Mach number for omega};
thus, the massless branch $\Omega_-$ is feasible, whereas the cyclotron branch $\Omega_+$ is unstable.
According to Ref.~\cite{vollhardt2013superfluid}, the speed of the first sound in fermionic superfluids is $c_s=v_F[\frac{1}{3}(1+F^s_0)(1+F^s_1/3)]^{1/2}$ 
with the Landau parameters $F^s_{0,1}$ and the Fermi velocity $v_F$.
As a result, one obtains
\begin{equation*}
    \tilde{c}_s^F
    =\frac{c_s^F}{\sigma/\tau}
    \sim\xi n^{\frac{1}{3}}\sqrt{\frac{m_v}{\pi\rho\xi^2}},
\end{equation*}
where $n$ denotes the particle number density.
The rescaled sound speed $\tilde{c}_s^F$ of the Fermi systems has an extra factor $\xi n^{\frac{1}{3}}\sim\tilde{c}_s^F/\tilde{c}_s^B$ compared with Bose systems.
This factor could be significantly greater than unity.
For example, for the superfluid phase of $^3$He-B, where the vortex dynamics was typically described by the vortex filament model, this factor becomes substantially large, yielding a large factor $\tilde{c}_s^F/\tilde{c}_s^B\gtrsim10^2$
~\footnote{
Using $\xi\approx10^3-10^4 \mathrm{\AA}$ and $n\sim10^{-27} \mathrm{m}^{-3}$ in the superfluid phase of $^3$He-B at zero temperature, we obtain $\tilde{c}_s^F\gg\tilde{c}_s^B$.
}.
Although the range with small $M$ for bosonic superfluids is narrow as shown in Fig.~\ref{fig:angular frequency},
that for fermionic superfluids can be substantially wide due to the large factor 
and then both branches can be stabilized.
For superfluid Fermi gases, however, 
this factor is on the order of unity in typical experiments, 
and the cyclotron branch $\Omega_+$ is unstable.
According to the Ginzburg--Landau theory,
the factor may be large when the healing length $\xi$ increases as the temperature approaches the superfluid critical temperature.
Thermal dissipation plays an important role in vortex dynamics, and we need to introduce a mutual friction force into our model.
This problem is beyond the scope of the current study and is an interesting topic for future work.
In the following analysis, we focus on the massless branch $\Omega_-$ as the stable frequency for bosonic and fermionic superfluids.

Finally, we investigate the circular orbit with $\Omega_-$.
Proving $\Omega_-(r)>\Omega_0(r)$ for $r\geq2\sigma$ can be done in a straightforward manner by using Eqs.~\eqref{eq:omega0} and \eqref{eq:omega}.
This relation indicates that the massless vortices are always delayed compared with the massive vortices along the same circular orbit.
We can then analytically evaluate the delay of the massless vortices by calculating the angle shift $\Delta\theta$ after period $2\pi/\Omega_0$;
\begin{equation}
    \Delta\theta_{\mathrm{ana}}(r)
    =2\pi\left(\frac{\Omega_-(r)}{\Omega_0(r)}-1\right).
    \label{eq:analytical angle shift}
\end{equation}
Both the angle shift $\Delta\theta_{\mathrm{ana}}(r)$ and the arc-length difference $\frac{r}{2}\Delta\theta_{\mathrm{ana}}(r)$ are monotonically decreasing functions of $r$.
To make $\frac{r}{2}\Delta\theta_{\mathrm{ana}}$ larger than $\sigma$, $2\sigma$, and $3\sigma$, restrictions $r\leq3.70\sigma$, $r\leq2.48\sigma$, and $r\leq2.17\sigma$ are needed, respectively, 
where the vortex mass length $\sigma=(m_v/\pi\rho\xi^2)^{1/2}\xi\lesssim\xi$ typically corresponds to the core size in our parameter range~\eqref{eq:range of vortex mass}.

\subsection{\label{sec:angle shift}Angle-shift enhancement}
To verify the vortex mass more effectively, the angle shift $\Delta\theta$ can be enhanced by changing the initial angular velocity $\dot{\theta}(t=0)$.
We assume that the initial velocity is manipulated by introducing a pinning potential, as demonstrated experimentally~\cite{kwon2021sound,samson2016deterministic}.
Although we can theoretically select an arbitrary initial velocity within the range of Eq.~\eqref{eq:Mach number}. 
The possible experimental range of the initial velocity must be considered.
The initial velocity is assigned to the vortex by capturing it with vortex pinning and operating it in a sophisticated manner.
If the speed of the pinning potential relative to the background superfluid velocity is higher than a certain threshold, the initial velocity of the vortex cannot be precisely controlled.
We apply the criteria described in Ref.~\cite{stockdale2021dynamical}, 
where the vortex is pinned when the relative speed is less than approximately $0.2c_s$.
This condition is represented by the Mach number defined in Eq.~\eqref{eq:Mach number for omega}, with $\Omega$ replaced by $\dot{\theta}(t=0)$ in $M$ as $\frac{r_0}{2c_s}|\dot{\theta}(0)-\Omega_0(r_0)|\lesssim0.2$ with $r_0\equiv r(t=0)$.
Correspondingly, the possible range of the initial angular velocity with $\dot{r}(t=0)=0$ is $\dot{\theta}_0^{(-)}\lesssim\dot{\theta}(0)\lesssim\dot{\theta}_0^{(+)}$ with 
\begin{equation}
    \dot{\theta}_0^{(\pm)}(r_0)
    \equiv\Omega_0(r_0)\pm\frac{0.2c_s}{r_0/2}.
    \label{eq:initial angular velocity}
\end{equation}

\begin{figure}[htbp]
\centering
\includegraphics[width=8.45cm]{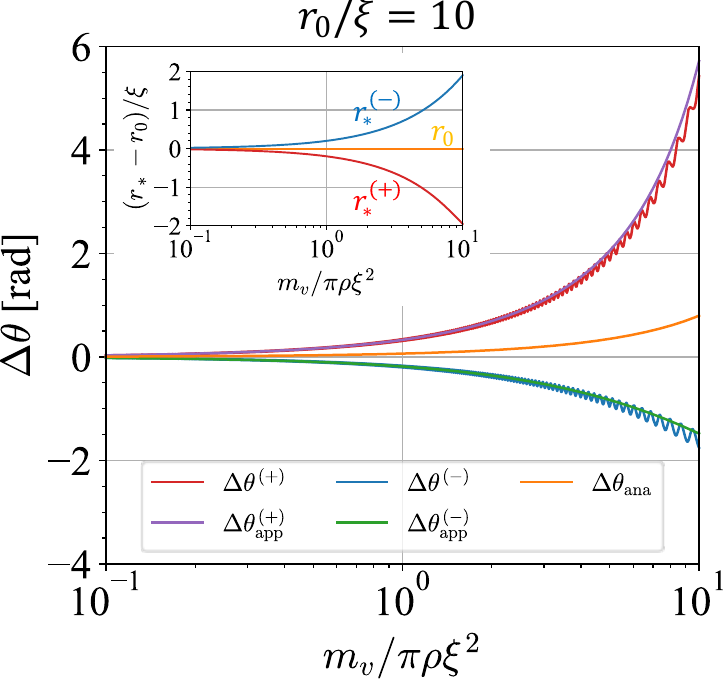}
\caption{
Plots of the angle shift $\Delta\theta$ after a period $2\pi/\Omega_0(r_0)$ with several initial angular velocities for $r_0/\xi=10$.
The graphs of $\Delta\theta^{(\pm)}$ are obtained numerically with $\dot{\theta}(0)=\dot{\theta}_0^{(\pm)}(r_0)$.
The graphs of $\Delta\theta^{(\pm)}_{\mathrm{app}}$ and $\Delta\theta_{\mathrm{ana}}$ are obtained approximately [see Eq.~\eqref{eq:approximate angle shift}] and analytically [see Eq.~\eqref{eq:analytical angle shift}].
The inset shows the analytical solutions of the difference $r_*-r_0$ rescaled with $\xi$.
The red, blue, and orange curves represent the results for $r_*=r_*^{(+)}$, $r_*^{(-)}$, and $r_0$ with $\dot{\theta}(0)=\dot{\theta}_0^{(+)}(r_0)$, $\dot{\theta}_0^{(-)}(r_0)$, and $\Omega_0(r_0)$, respectively.
}
\label{fig:angle_r0_10}
\end{figure}

\begin{figure}[tbp]
\centering
\includegraphics[width=8.45cm]{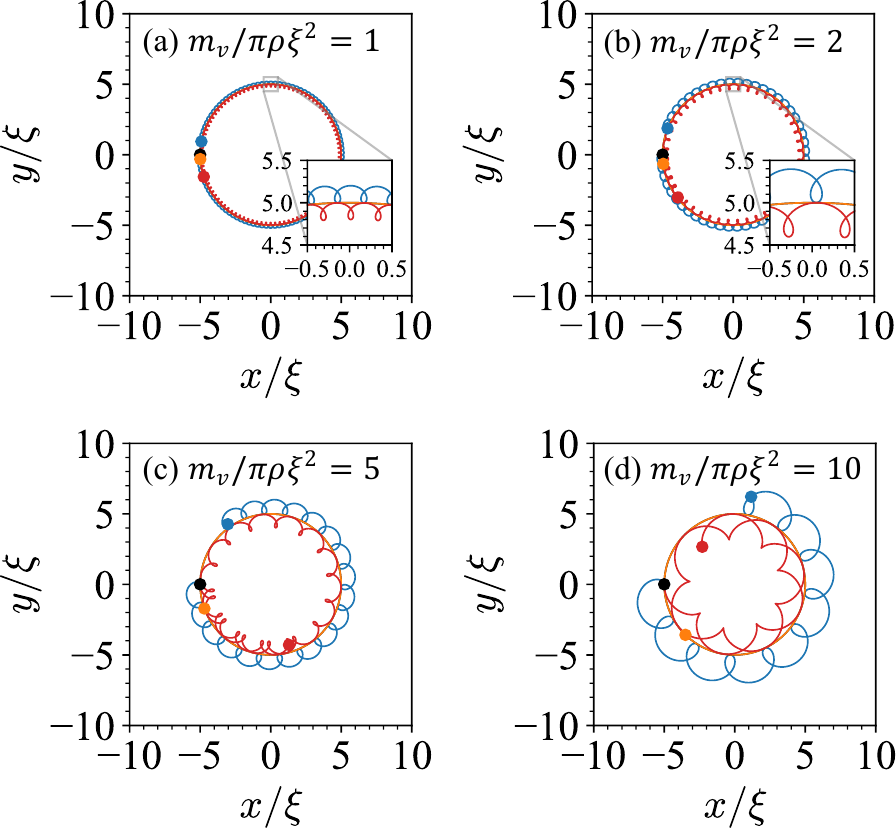}
\caption{
Typical trajectories of the first vortex at $\bm{r}_1=-\bm{r}_R/2=(-\frac{r}{2}\cos\theta,-\frac{r}{2}\sin\theta)$ with (a)$m_v/\pi\rho\xi^2=1$, (b)$m_v/\pi\rho\xi^2=2$, (c)$m_v/\pi\rho\xi^2=5$, and (d)$m_v/\pi\rho\xi^2=10$.
The red, blue, and orange lines represent the trajectories with the initial conditions $(r(0),\theta(0),\dot{r}(0),\dot{\theta}(0))=(r_0,0,0,\dot{\theta}_0^{(+)}(r_0))$, $(r_0,0,0,\dot{\theta}_0^{(-)}(r_0))$, and $(r_0,0,0,\Omega_-(r_0))$ with $r_0=10\xi$, respectively.
Black lines represent the trajectories of the massless vortices given by $\bm{r}_1(t)=(-\frac{r_0}{2}\cos\Omega_0(r_0)t,-\frac{r_0}{2}\sin\Omega_0(r_0)t)$ in common with (a)-(d).
In (a) and (b), the insets show the closeups of trajectories to confirm the radial oscillation.
Each marker shows the position of the first vortex at $t=2\pi/\Omega_0(r_0)$, and its size corresponds to the vortex core size.
The position of the second vortex is $\bm{r}_2=\bm{r}_R/2=-\bm{r}_1$, which is not plotted for convenience.
}
\label{fig:trajectory}
\end{figure}

In Fig.~\ref{fig:angle_r0_10}, we plot the angle shifts $\Delta \theta^{(\pm)}$ with $\dot{\theta}(0)=\dot{\theta}_0^{(\pm)}$ is a function of $m_v/\pi\rho\xi^2$ for $r_0/\xi=10$, which are obtained by the numerical simulation of the massive PVM
~\footnote{
We performed numerical simulations of the massive PVM using the velocity Verlet algorithm.
In all the simulations, we numerically solved Eq.~\eqref{eq:dimensionless EOM} with rescaled time $\tilde{t}=t/\tau$ and length $\tilde{\bm{r}}_i=\bm{r}_i/\sigma$, and set the time step $\Delta\tilde{t}\leq0.001$.
}.
Assuming experimental observations of the trajectories of a vortex--vortex pair, we use 
\begin{equation}
    \tau_\xi
    =\frac{2\pi\xi^2}{\kappa}
    \label{eq:time scale, xi}
\end{equation} 
and $\xi$ as time and length units instead of $\tau$ and $\sigma$
~\footnote{
By rescaling the function of $r$, such as Eq.~\eqref{eq:analytical angle shift}, with $\tau_\xi$ and $\xi$ instead of $\tau$ and $\sigma$, 
new factors $\tau/\tau_\xi=\frac{1}{2}(m_v/\pi\rho\xi^2)$ and $\sigma/\xi=(m_v/\pi\rho\xi^2)^{1/2}$ appear.Accordingly, this function has two variables, $r/\xi$ and $m_v/\pi\rho\xi^2$.
Note that $m_v/\pi\rho\xi^2$ is an unknown quantity while $r/\xi$ is experimentally operable.
}.
%where $\xi$ corresponds to the observation resolution.
Evidently, $|\Delta\theta^{(\pm)}|$ monotonically increases, and the difference $\Delta\theta^{(+)}-\Delta\theta^{(-)}$ is larger than $\Delta\theta_{\mathrm{ana}}$ for any $m_v/\pi\rho\xi^2$.
Thus, by controlling the initial velocity, the impact of the vortex mass on the angle shift can be enhanced, and the arc-length difference $\frac{r_0}{2}(\Delta\theta^{(+)}-\Delta\theta^{(-)})$ is large enough to be more observable than $\frac{r_0}{2}\Delta\theta_{\mathrm{ana}}$.
Here, by regarding the motion of massive vortices as the circular motion with $r_*$, 
which is the equilibrium point of the effective radial potential $V_{\mathrm{eff}}$, 
we obtain approximate solutions of $\Delta\theta^{(\pm)}$ as a function of $r_0/\xi$ and $m_v/\pi\rho\xi^2$: 
\begin{align}
    \Delta\theta_{\mathrm{app}}^{(\pm)}
    &=2\pi\left(\frac{\Omega_-(r_*^{(\pm)})}{\Omega_0(r_0)}-1\right) \notag \\
    &=2\pi\left\{\frac{1}{2}\frac{(r_0/\xi)^2}{m_v/\pi\rho\xi^2}
      \left[1-\sqrt{1-4\frac{m_v/\pi\rho\xi^2}{(r_*^{(\pm)}/\xi)^2}}\right]-1\right\}
    \label{eq:approximate angle shift}
\end{align}
with
\begin{align}
    \frac{r_*^{(\pm)}}{\xi}
    &\equiv \frac{r_*(p_\theta=p_\theta^{(\pm)})}{\xi} \notag \\ 
    &=\sqrt{2\frac{m_v}{\pi\rho\xi^2}+\sqrt{4\left(\frac{m_v}{\pi\rho\xi^2}\right)^2+\left(\frac{4p_\theta^{(\pm)}}{\rho\kappa\xi^2}\right)^2}}
    \label{eq:local minimum plus-minus}, \\
    \frac{4p_\theta^{(\pm)}}{\rho\kappa\xi^2}
    &\equiv\frac{4}{\rho\kappa\xi^2}\left(\frac{m_v}{2}r_0^2\dot{\theta}_0^{(\pm)}-\frac{\rho\kappa}{4}r^2_0\right) \notag \\
    &=\frac{m_v}{\pi\rho\xi^2}\left(\frac{r_0}{\xi}\right)^2\tau_\xi\dot{\theta}_0^{(\pm)}-\left(\frac{r_0}{\xi}\right)^2,
    \label{eq:azimuth momentum plus-minus}
\end{align}
and 
\begin{align}
    \tau_\xi\dot{\theta}_0^{(\pm)}
    &=\frac{2}{(r_0/\xi)^2}\pm\frac{0.4}{r_0/\xi}, 
    \label{eq:scaled initial angular velocity}
\end{align}
where Eqs.~\eqref{eq:omega0}, \eqref{eq:azimuth momentum}, \eqref{eq:local minimum}, \eqref{eq:omega}, \eqref{eq:initial angular velocity}, \eqref{eq:time scale, xi}, and $c_s=\kappa/2\pi\xi=\xi/\tau_\xi$ are used.
As shown in Fig.~\ref{fig:angle_r0_10}, $\Delta\theta_{\mathrm{app}}^{(\pm)}$ agrees very well with the numerical solutions.
This ensures that the vortex mass can be estimated quantitatively by observing $\Delta\theta^{(\pm)}$ and comparing it with $\Delta\theta_{\mathrm{app}}^{(\pm)}$.
Therefore, the angular shift can be used as a tester for different predictions of the vortex mass,
and Fig.~\ref{fig:angle_r0_10} and the set of equations~\eqref{eq:approximate angle shift}, \eqref{eq:local minimum plus-minus}, \eqref{eq:azimuth momentum plus-minus}, and \eqref{eq:scaled initial angular velocity}  provide sufficient information to estimate the vortex mass efficiently and precisely through experimental observations.

\begin{figure}[tbp]
\centering
\includegraphics[width=8.45cm]{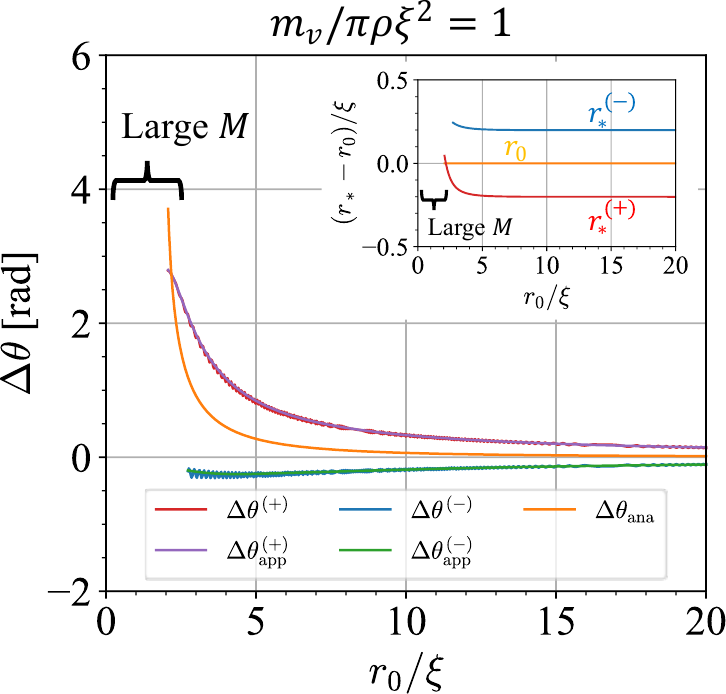}
\caption{
Plots of the angle shift $\Delta\theta$ after a period $2\pi/\Omega_0(r_0)$ with several initial angular velocities for $m_v/\pi\rho\xi^2=1$.
The graphs of $\Delta\theta^{(\pm)}$ are obtained numerically with $\dot{\theta}(0)=\dot{\theta}_0^{(\pm)}(r_0)$.
The graphs of $\Delta\theta^{(\pm)}_{\mathrm{app}}$ and $\Delta\theta_{\mathrm{ana}}$ are obtained approximately [see Eq.~\eqref{eq:approximate angle shift}] and analytically [see Eq.~\eqref{eq:analytical angle shift}].
The inset shows the analytical solutions of the difference $r_*-r_0$ rescaled with $\xi$.
The red, blue, and orange curves represent the results for $r_*=r_*^{(+)}$, $r_*^{(-)}$, and $r_0$ with $\dot{\theta}(0)=\dot{\theta}_0^{(+)}(r_0)$, $\dot{\theta}_0^{(-)}(r_0)$, and $\Omega_0(r_0)$, respectively.
Each graph is cut off for small $m_v/\pi\rho\xi^2$, where the voretex velocity is beyond the applicable condition of the massive PVM, $M\leq0.3$, with $c_s=\kappa/2\pi\xi$.
}
\label{fig:angle_m_1}
\end{figure}

As shown in Fig.~\ref{fig:angle_r0_10}, we find the oscillation of the graphs of $\Delta\theta^{(\pm)}$, especially for a large $m_v/\pi\rho\xi^2$, which is caused by the oscillation of the vortices in the radial direction.
Because the pseudo-momentum $p_\theta$ changes according to the initial angular velocity, the initial distance $r_0$ is not the minimum $r_*$ of the effective radial potential $V_{\mathrm{eff}}$ except for the circular orbit solutions.
Thus, even if the initial radial velocity $\dot{r}(0)$ is zero, the trajectory is not necessarily circular.
Some trajectories of the first vortex for several values of $m_v/\pi\rho\xi^2$ after a period $2\pi/\Omega_0(r_0)$ are shown in Fig.~\ref{fig:trajectory}.
Clearly, the amplitude of the oscillation becomes larger as $m_v/\pi\rho\xi^2$ increases.
The amplitude can be estimated analytically from the difference between $r_*$ and $r_0$, as shown in the inset of Fig.~\ref{fig:angle_r0_10}.
If the amplitude is significantly larger than the healing length $\xi$,
the vortex mass can be verified by observing the oscillation instead of the angular shift.
However, it is difficult to observe the amplitude within the investigated parameter range [see Fig.~\ref{fig:trajectory}(a)(b)].
Therefore, observing the difference $\Delta\theta^{(+)}-\Delta\theta^{(-)}$ is much more useful than the radial oscillation for efficiently detecting the impact of the vortex mass.

For reference, we also show graphs of $\Delta\theta^{(\pm)}$, $\Delta\theta_{\mathrm{app}}^{(\pm)}$, $\Delta\theta_{\mathrm{ana}}$, and $r_*^{(\pm)}-r_0$ as functions of $\frac{r_0}{\xi}$ in Fig.~\ref{fig:angle_m_1}.
The approximate solutions $\Delta\theta_{\mathrm{app}}^{(\pm)}$ also agree very well with the numerical solutions $\Delta\theta^{(\pm)}$ because $|r_*^{(\pm)}-r_0|$ is much smaller than $\xi$.
Contrary to Fig.~\ref{fig:angle_r0_10}, $\Delta\theta^{(+)}-\Delta\theta^{(-)}$ and $\Delta\theta_{\mathrm{ana}}$ decrease monotonically, and each graph is cut off because of the condition $M\leq0.3$.
Thus, a minimum value of $r_0$ restricts the experimentally operable range of $r_0$.

\subsection{\label{sec:instability} A cause of the splitting instability}
The vortex mass is related to the splitting instability of a doubly quantized vortex in a uniform superfluid at absolute zero \cite{takeuchi2018doubly}.
The lowest radius of the circular orbit solutions in Eq.~\eqref{eq:local minimum} implies that the vortex mass causes dynamic instability.
A pair of quantum vortices with the same circulation $\kappa$ cannot maintain their distance $r(<2\sigma)$ and are dynamically unstable.
This is in clear contrast to the fact that the two vortices can assume a circular orbit at any distance in a conventional massless PVM.

During the splitting instability process, energy is emitted to infinity through the spontaneous radiation of helical phonons~\cite{takeuchi2018doubly}.
Assuming that the energetics of this system is divided into the incompressible component described in the PVM and the compressible one consisting of phonons and the total energy is conserved, 
the phonon emission causes energy dissipation in the former.
Dissipation in the PVM system can affect the frictional force of the relative vortex motion in the radial direction.
Thus, the final state of the splitting instability is a circular orbit solution with an equilibrium distance $r_*(p_\theta)$, as discussed in Sec.~\ref{sec:circular orbits}.
Considering that this is a dissipation process from the closest state of $r\to 0$, the final state of the splitting instability can be evaluated based on the result for $p_\theta = 0$;
the circular orbit solution with distance $r_{\rm fin} \gtrsim r_*(p_\theta=0)=2\sigma$.

In other words, the vortex mass is calculated from the final state resulting from splitting instability.
By taking $\xi$ as the length unit, 
the vortex mass $m_v$ is estimated with the final distance $r_{\rm fin}$ as follows:
\begin{equation}
    \frac{m_v}{\pi\rho\xi^2}\lesssim\left(\frac{r_{\rm fin}}{2\xi}\right)^2.
    \label{eq:upper limit of vortex mass}
\end{equation}
This expression enables us to estimate the upper limit of the vortex mass from the observation of the final state of the splitting instability.

An order estimation from a previous experiment on superfluid Bose gases with $r_{\rm fin}\sim 50~{\rm \mu m}$ and $\xi\sim 0.4~{\rm \mu m}$~\cite{shin2004dynamical} yielded $m_v/\pi\rho\xi^2\lesssim 10^3$.
This result seems too large to be supported by existing theoretical models.
However, drawing conclusions from this discrepancy is premature.
This is because the condensate was trapped by a harmonic potential in the above experiment, and the observed instability could be strongly influenced by three-dimensional nonuniformity~\cite{isoshima2008vortex}.
On the other hand, no clear splitting was confirmed in the mean-field model of dilute Bose gases at absolute zero with numerical simulations of the Gross--Pitaevskii equation~\cite{aranson1996stability}.
This implies that the vortex mass is very small or negligible in a scalar bosonic superfluid at the mean-field level~\cite{groszek2018motion}.
Interestingly, according to low-energy effective-field theory for superfluid Fermi gases~\cite{klimin2015finite,klimin2016finite,klimin2018diversified}, the splitting instability may drastically change its behavior in the BEC--BCS crossover~\cite{van2024splitting}.
In particular, they observed a significant vortex splitting on the BCS side, although the instability remains weak on the BEC side.
This suggests that the impact of vortex mass is more significant in fermionic superfluids.

Furthermore, the previous research~\cite{Patrick2023stability} shows that multiply quantized vortices can be stabilized by adding a small number of atoms of a second species to the vortex cores in immiscible binary BECs.
The mean-field potential due to the population of the component inside the core is crucial to the stabilization in the study
while the structure inside the vortex core is neglected in our work.
According to Ref.~\cite{hayashi2013instability}, the stability is microscopically described by ripplons living on the interface between the two components.
Therefore, our theory with the PVM cannot describe such an initial process of the instability.
Rather, we explain the cause of separating the two singly-quantized vortices in the later stage where their vortex cores do not overlap.

\section{\label{sec:vortex-antivortex}A vortex--antivortex pair}
We analyze the dynamics of a pair of quantum vortices with opposite circulations $\Gamma_1=-\Gamma_2=\kappa$.
The problem is formulated in Sec.~\ref{sec:effective radial potential} and we investigate the translational trajectory as the simplest case in Sec.~\ref{sec:translational trajectory}.
We consider manipulating the initial velocity as described in the previous section and demonstrate that the vortex mass leads to pair annihilation in Sec.~\ref{sec:relative oscillation and collision}.

\subsection{\label{sec:effective relative potential}Effective relative potential}
For the problem of a vortex--antivortex pair, the equation of motion with respect to $\bm{r}_1$ ($\bm{r}_2$) is given by Eq.~\eqref{eq:dimensionless EOM} with the plus (minus) sign of the first term and the minus sign of the second term.
In the same manner as that described in Sec.~\ref{sec:effective radial potential}, we transform the equations into
\begin{align}
    M_v\frac{d^2\bm{r}_G}{dt^2} 
    &=-\rho\kappa\bm{e}_z\times\frac{d\bm{r}_R}{dt},
    \label{eq:EOM for vortex-antivortex,G}\\
    \mu_v\frac{d^2\bm{r}_R}{dt^2} 
    &=-\rho\kappa\bm{e}_z\times\frac{d\bm{r}_G}{dt}-\frac{\rho\kappa^2}{2\pi}\frac{\bm{r}_R}{|\bm{r}_R|^2}
    \label{eq:EOM for vortex-antivortex,R}
\end{align}
with $\bm{r}_G=(\bm{r}_1+\bm{r}_2)/2$, $\bm{r}_R=\bm{r}_2-\bm{r}_1$, $M_v=2m_v$, and $\mu_v=m_v/2$.
Unlike the vortex--vortex pair [Eqs.~\eqref{eq:EOM for vortex-vortex,G} and \eqref{eq:EOM for vortex-vortex,R}], the equations with respect to $\bm{r}_G$ and $\bm{r}_R$ are not independent.

To clarify this analysis, we introduce the Hamiltonian of the two-body vortex problem.
According to Appendix~\ref{app:canonical transformation}, the canonical momenta are given by 
\begin{align}
    \bm{p}_G
    &=M_v\dot{\bm{r}}_G+\rho\kappa\bm{e}_z\times\bm{r}_R
    \label{eq:momentum_G}, \\
    \bm{p}_R
    &=\mu_v\dot{\bm{r}}_R.
    \label{eq:momentum_R}
\end{align}
The pseudomomentum $\bm{p}_G$ is conserved because $\dot{\bm{p}}_G=\partial H/\partial\bm{r}_G=0$.
Consequently, the Hamiltonian expressed in Eq.~\eqref{eq:Hamiltonian for vortex-antivortex, appendix} is  
\begin{equation}
    H(\bm{r}_G,\bm{r}_R,\bm{p}_G,\bm{p}_R)
    =\frac{\bm{p}_R^2}{2\mu_v}
    +U_{\mathrm{eff}}(\bm{r}_R,\bm{p}_G)
    \label{eq:Hamiltonian for vortex-antivortex}
\end{equation}
with effective potential
\begin{equation}
    U_{\mathrm{eff}}(\bm{r}_R,\bm{p}_G)\\
    =\frac{\rho^2\kappa^2}{8\mu_v}\left(\bm{r}_R+\bm{e}_z\times\frac{\bm{p}_G}{\rho\kappa}\right)^2
    +\frac{\rho\kappa^2}{2\pi}\ln\frac{|\bm{r}_R|}{\sigma},
    \label{eq:effective relative potential}
\end{equation}
where a constant is added and included in the last term.
The first term on the right side of Eq.~\eqref{eq:Hamiltonian for vortex-antivortex} is the relative kinetic energy.
The first term on the right side of Eq.~\eqref{eq:effective relative potential} denotes the harmonic potential associated with the pseudomomentum.
The second term represents the logarithmic interaction potential.
Since $\bm{p}_G$ is conserved,
we can treat it as a parameter, and $U_{\mathrm{eff}}(\bm{r}_R)$ is called the effective relative potential.
Eventually, we succeed in decoupling the center of mass and relative motions.

To describe the relative motion more clearly, we consider the symmetrical initial condition with respect to the $y$ axis:
\begin{align*}
    \bm{r}_R(0)
    &=(x_R(0),0) \quad (x_R(0)>0), \\ 
    \dot{\bm{r}}_G(0)
    &=(0,v_G(0))
\end{align*}
As the equilibrium point of the harmonic potential in Eq.~\eqref{eq:effective relative potential},
we define $\bar{\bm{r}}_R\equiv-\bm{e}_z\times\frac{\bm{p}_G}{\rho\kappa}=(\bar{x}_R,\ \bar{y}_R)$ as
\begin{equation}
    \begin{split}
    \bar{x}_R
    &\equiv\tilde{\bar{x}}_R\sigma
    =\left(1+2\frac{v_G(0)\tau}{x_R(0)}\right)x_R(0),\\
    \bar{y}_R
    &=0,
    \end{split}
    \label{eq:xbar}
\end{equation}
where Eq.~\eqref{eq:momentum_G} is used.
Then, $U_{\mathrm{eff}}(\bm{r}_R)$ has an extremum at $\bm{r}_R=(x_{R*}, y_{R*})$ with 
\begin{equation}
    \begin{split}
    x_{R*}
    &\equiv\tilde{x}_{R*}\sigma
    =\frac{\bar{x}_R}{2}\left(1\pm\sqrt{1-\frac{4}{(\bar{x}_R/\sigma)^2}}\right) \quad 
    (\bar{x}_R\geq2\sigma),\\
    y_{R*}
    &=0.
    \end{split}
    \label{eq:extremum}
\end{equation}
As $\partial^2U_{\mathrm{eff}}(\bm{r}_R)/\partial y_R^2>0$ holds for any $x_R$ and $y_R$,
the $y$-component of the relative motion is determined to be $y_R(t)=y_{R*}=0$ if $y_R(t=0)=0$ and $\dot{y}_R(t=0)=0$.
For simplicity, we set $y_R(t)=0$ and choose three values, $x_R(0)$, $\dot{x}_R(0)$, and $v_G(0)$, as the initial conditions.
Therefore, the relative motion is reduced to one-dimensional motion in the effective relative potential $U_{\mathrm{eff}}(x_R)$.
Finally, we obtain the rescaled form
\begin{equation*}
    \tilde{U}_{\mathrm{eff}}
    \equiv\frac{U_{\mathrm{eff}}(x_R)}{m_v(\sigma/\tau)^2}
    =\frac{1}{4}(\tilde{x}_R-\tilde{\bar{x}}_R)^2
    +\frac{1}{2}\ln|\tilde{x}_R|.
\end{equation*}

\begin{figure}[tbp]
\centering
\includegraphics[width=8.45cm]{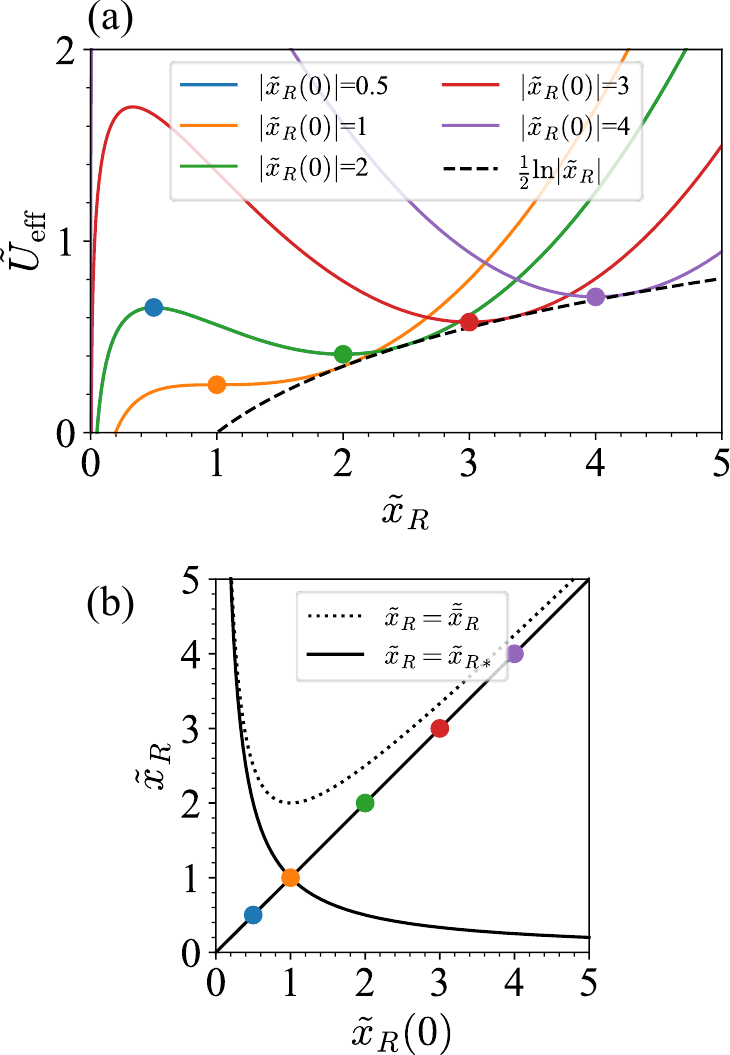}
\caption{
(a) Plots of the effective relative potential $\tilde{U}_{\mathrm{eff}}$ (solid) when $v_G(0)=V_0(x_R(0))$ for several values of $\tilde{x}_R(0)$.
The dotted guideline shows the logarithmic interaction potential.
(b) The local maximum and minimum points $\tilde{x}_{R*}$ given by Eq.~\eqref{eq:extremum} (solid) and the equilibrium point $\tilde{\bar{x}}_R$ of the harmonic potential given by Eq.~\eqref{eq:equilibrium point} (dotted).
In each panel, markers correspond to the initial position $\tilde{x}_R(0)$ of the vortex.
}
\label{fig:potential for vortex-antivortex}
\end{figure}

\subsection{\label{sec:translational trajectory}Translational trajectory solution}
As a simple initial condition, we assume that the center of mass velocity is equal to the velocity of the massless vortices given by Eq.~\eqref{eq:V0}; 
\begin{equation*}
    v_G(0)=V_0(x_R(0)).
\end{equation*}
Inserting it into Eq.~\eqref{eq:xbar}, we obtain
\begin{equation}
    \bar{x}_R=x_R(0)+\frac{\sigma^2}{x_R(0)}.
    \label{eq:equilibrium point}
\end{equation}
Consequently, the specific forms of $U_{\mathrm{eff}}$ and $x_{R*}$ are determined and shown in Fig.~\ref{fig:potential for vortex-antivortex}.
The inflection point corresponds to the initial position $x_R(0)$ when $x_R(0)=\sigma$.
The local minimum (maximum) point corresponds to $x_R(0)$ when $x_R(0)>\sigma$ [$x_R(0)<\sigma$].
Therefore, when the relative kinetic energy is zero at $t=0$ with $\dot{x}_R(0)=0$, the intervortex distance becomes $x_R(t)=x_R(0)$. 
Accordingly, Eq.~\eqref{eq:EOM for vortex-antivortex,G} shows that the acceleration term becomes zero, and $v_G(t)=v_G(0)=V_0(x_R(0))$ holds.
Thus, the dynamics of a vortex--antivortex pair are translational motion, which is identical to that of massless vortices.

\begin{figure}[tbp]
\centering
\includegraphics[width=8.45cm]{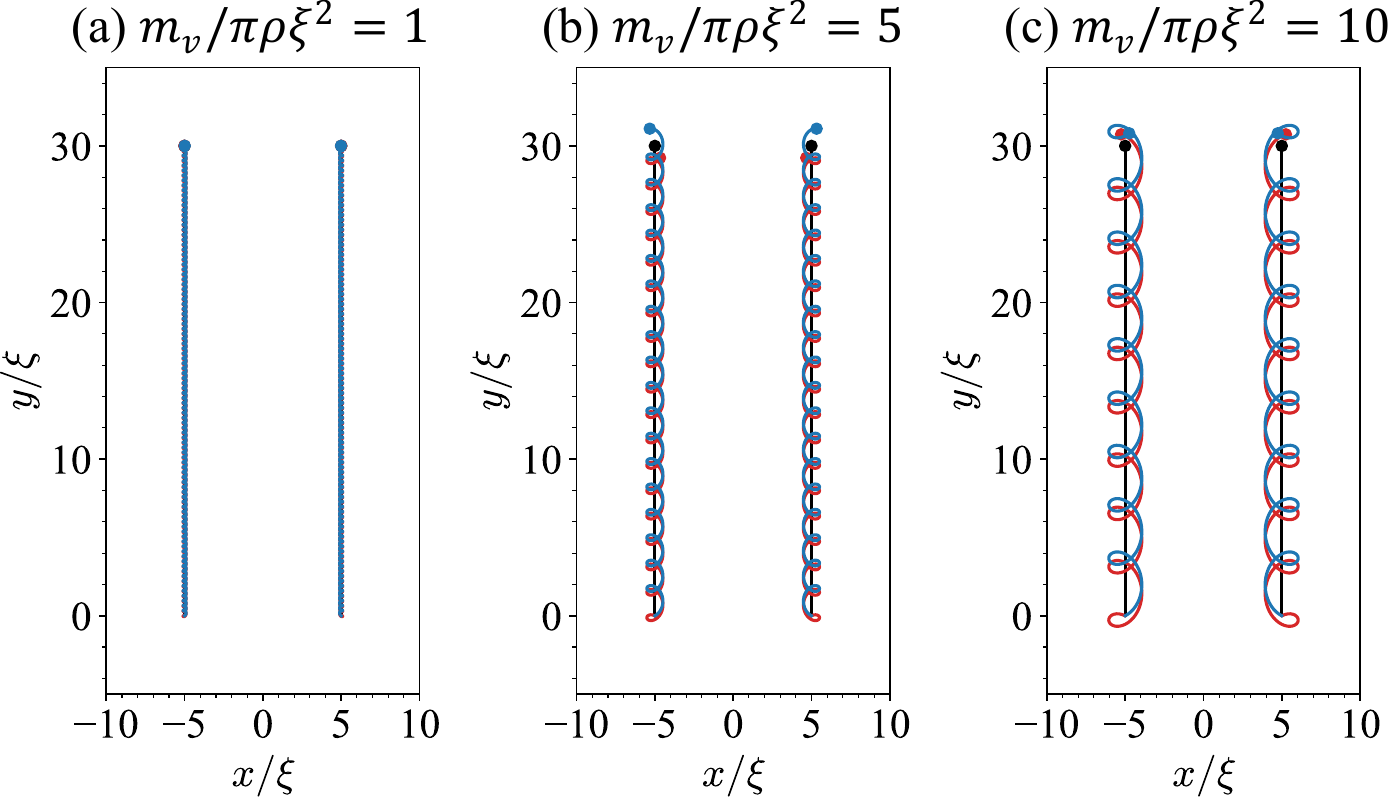}
\caption{
Typical trajectories of the first and second vortices at $\bm{r}_1=\bm{r}_G-\bm{r}_R/2$ and $\bm{r}_2=\bm{r}_G+\bm{r}_R/2$ with (a)$m_v/\pi\rho\xi^2=1$, (b)$m_v/\pi\rho\xi^2=5$, and (c)$m_v/\pi\rho\xi^2=10$.
The red and blue lines represent the trajectories with the initial conditions $\bm{r}_R(0)=(x_R(0),0)$, $\dot{\bm{r}}_R(0)=(\pm0.4c_s,0)$, and $\bm{v}_G(0)=(0,V_0(x_R(0)))$ with $x_R(0)=10\xi$, respectively.
Black lines represent the trajectories of massless vortices given by $\bm{r}_1(t)=(-x_R(0)/2,V_0(x_R(0))t)$ and $\bm{r}_2(t)=(+x_R(0)/2,V_0(x_R(0))t)$ in common with (a)-(c).
Each marker shows the position of the vortex at $t=300\tau_\xi$, and its size corresponds to the vortex core size.
Note that the three different trajectories overlap in (a).
}
\label{fig:verification1}
\end{figure}

However, when the relative kinetic energy is nonzero with $\dot{x}_R(0)\neq0$, oscillations in the $x$-axis direction occur (see Fig.~\ref{fig:verification1}).
The amplitude of the oscillation is obtained by calculating the turning points from the conservation of total energy with Eq.~\eqref{eq:Hamiltonian for vortex-antivortex}.
In the next section, we analytically investigate the amplitude depending on the initial velocity.

\begin{figure}[htbp]
\centering
\includegraphics[width=8.45cm]{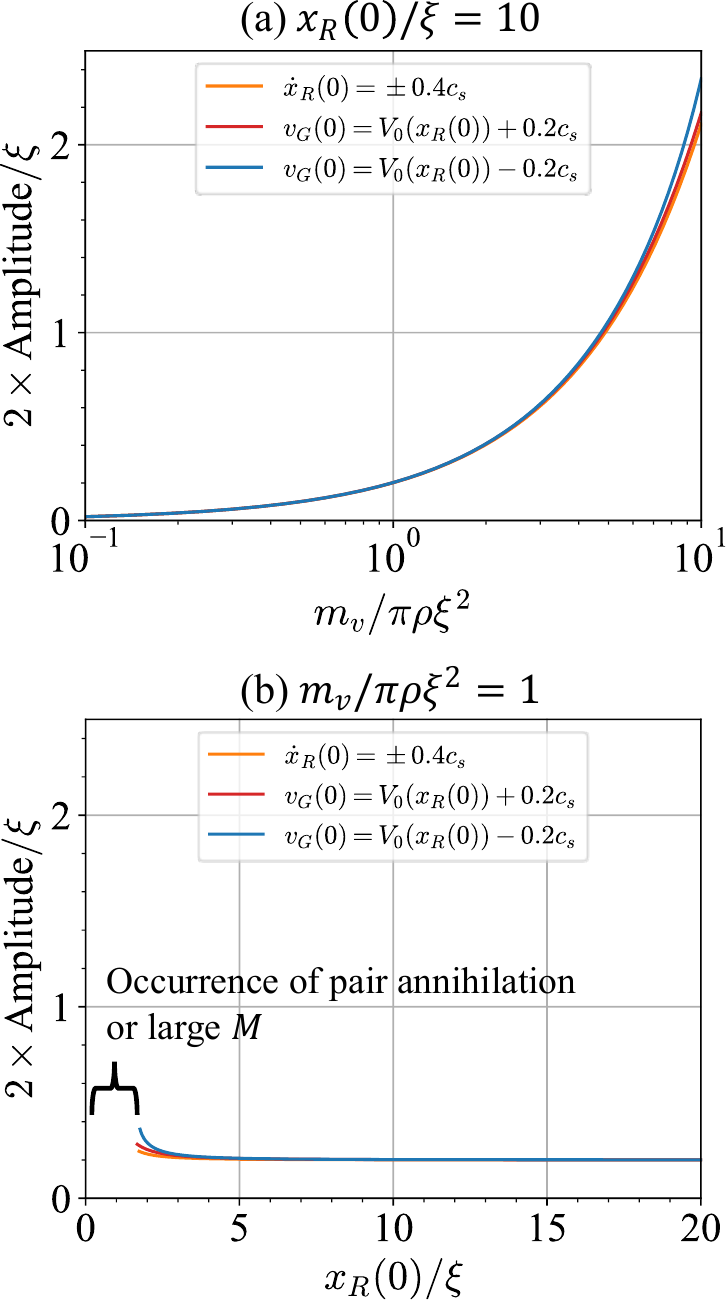}
\caption{
Plots of the amplitude of the relative oscillation rescaled with $\xi$ as a function of (a) $m_v/\pi\rho\xi^2$ and (b) $x_R(0)/\xi$.
In both panels, a fixed parameter is $x_R(0)/\xi=10$ and $m_v/\pi\rho\xi^2=1$, respectively.
The orange, red, and blue curves correspond to analytical results when $(\dot{x}_R(0),v_G(0))=(\pm0.4c_s,V_0(x_R(0))$, $(0,V_0(x_R(0))+0.2c_s)$, and $(0,V_0(x_R(0))-0.2c_s)$.
Each graph in (b) is cut off for small $m_v/\pi\rho\xi^2$, where pair annihilation of vortices occurs or the vortex velocity is beyond the applicable condition of the massive PVM, $M\leq0.3$, with $c_s=\kappa/2\pi\xi$.
}
\label{fig:amplitude}
\end{figure}

\subsection{\label{sec:relative oscillation and collision}Relative oscillation and collision} 
Similar to Sec.~\ref{sec:angle shift}, we consider manipulating the initial velocity to verify the vortex mass from the motion in the $x$-direction.
Here, we use $\tau_\xi$ and $\xi$ as time and length units,
respectively, assuming experimental observations of the amplitude.

As shown in the previous section, the intervortex distance $x_R(t)$ approaches zero if the relative energy is nonzero for $x_R(0)\leq\sigma$.
Then, owing to the attractive force caused by the logarithmic interaction in the presence of the vortex mass,
the two vortices collide and annihilate each other (see Fig.~\ref{fig:potential for vortex-antivortex}).
This description of the massive PVM shows the remarkable result that the pair annihilation of vortices can occur without a mutual friction and thermal fluctuations. 
Although the previous research~\cite{richaud2023mass-driven} pointed out that massive vortices can collide in immiscible binary BECs at zero temperature, 
they introduce a harmonic potential only for the minor component trapped in the core of vortices in the major component.
The force due to the potential obscured the intrinsic effect by the vortex mass.
In contrast, the effect predicted by our theory is purely intrinsic to the vortex mass without any external force.
In our case, pair annihilation typically occurs in the range $x_R(0)\leq\sigma=(m_v/\pi\rho\xi^2)^{1/2}\xi\lesssim\xi$
with Eq.~\eqref{eq:range of vortex mass},
where the cores of the two vortices overlap.
The vortex mass is then ill-defined;
thus, this initial distance is inappropriate for the massive PVM.
Hence, we attempt to verify the vortex mass in another manner.

\begin{figure}[tbp]
\centering
\includegraphics[width=8.45cm]{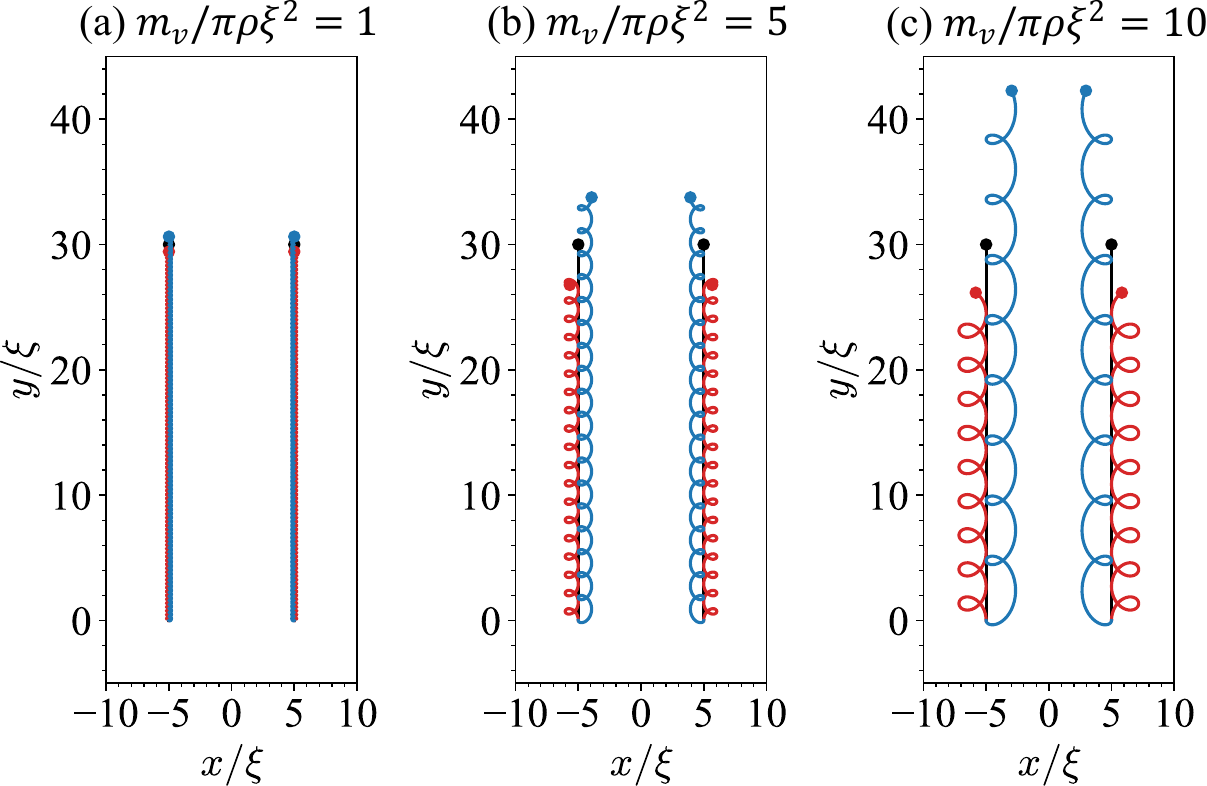}
\caption{
Typical trajectories of the first and second vortices at $\bm{r}_1=\bm{r}_G-\bm{r}_R/2$ and $\bm{r}_2=\bm{r}_G+\bm{r}_R/2$ with (a)$m_v/\pi\rho\xi^2=1$, (b)$m_v/\pi\rho\xi^2=5$, and (c)$m_v/\pi\rho\xi^2=10$.
The red and blue lines represent the trajectories with the initial conditions $\bm{r}_R(0)=(x_R(0),0)$, $\dot{\bm{r}}_R(0)=(0,0)$, and $\bm{v}_G(0)=(0,V_0(x_R(0))\pm0.2c_s)$ with $x_R(0)=10\xi$, respectively.
Black lines represent the trajectories of massless vortices given by $\bm{r}_1(t)=(-x_R(0)/2,V_0(x_R(0))t)$ and $\bm{r}_2(t)=(+x_R(0)/2,V_0(x_R(0))t)$ in common with (a)--(c) in both Fig.~\ref{fig:verification1} and this figure.
Each marker shows the position of the vortex at $t=300\tau_\xi$, and its size corresponds to the vortex core size.
Note that the three different trajectories overlap in (a).
}
\label{fig:verification2}
\end{figure}

We also consider controlling the initial velocity using vortex pinning.
When $v_G(0)=V_0(x_R(0))$, the maximum and minimum values of $\dot{x}_R(0)$ are
\begin{equation*}
    \dot{x}_R(0)=\pm0.4c_s
\end{equation*}
using $M=0.2$ (refer the discussion of the Mach number in Sec.~\ref{sec:angle shift}) with $u=|\dot{x}_R(0)/2|$ [see Eq.~\eqref{eq:relative motion speed}] and $c_s=\kappa/2\pi\xi$.
We show some vortex trajectories for several values of $m_v/\pi\rho\xi^2$ in Fig.~\ref{fig:verification1}.
As shown in the graph, the oscillation amplitude in the $x-$axis direction becomes larger as $m_v/\pi\rho\xi^2$ increases.
The amplitude is calculated analytically and plotted as a function of $m_v/\pi\rho\xi^2$ in Fig.~\ref{fig:amplitude}(a) and as a function of $x_R(0)/\xi$ in Fig.~\ref{fig:amplitude}(b).
The graphs (orange) in Fig.~\ref{fig:amplitude}(a) increase monotonically, and the variations of the graphs in Fig.~\ref{fig:amplitude}(b) are small.
For $m_v/\pi\rho\xi^2=1$, the amplitude is much smaller than the healing length $\xi$;
therefore, it may be difficult to observe.
Notably, the graphs in Fig.~\ref{fig:amplitude}(b) are cut off because of the applicable condition $M\leq0.3$ for the massive PVM.

Furthermore, instead of manipulating $\dot{x}_R(0)$, we consider different initial velocities; 
\begin{align*}
    \dot{x}_R(0)&=0, \\ 
    v_G(0)&=V_0(x_R(0))\pm0.2c_s,
\end{align*}
where $M=0.2$ holds.
In Fig.~\ref{fig:amplitude}, the graphs (red and blue) of the amplitude agree well with previous results (orange).
This indicates that it is also difficult to observe the amplitude, as in the above case.
However, as shown in Fig.~\ref{fig:verification2}, 
we can see that the massive vortices move behind (ahead of) massless voritces in the $y-$axis direction when $v_G(0)=V_0(x_R(0))+0.2c_s$ [$v_G(0)=V_0(x_R(0))-0.2c_s$].
The distance between the massive and massless vortices may be sufficiently large to be observed after a long time.
It would be an interesting topic for future work to suggest specific methods to observe the distance efficiently.

\section{\label{sec:conclusion}Summary and discussion}
We theoretically investigated the two-dimensional dynamics of quantum vortices with inertia based on the massive PVM.
The time and length scales in many-body problems of massive vortices are respectively characterized by
the vortex mass time $\tau$ and the vortex mass length $\sigma$,
determined by the circulation quantum, fluid density, and vortex mass [Eqs.~\eqref{eq:vortex mass time} and \eqref{eq:vortex mass length}].
Considering the circular motion of two like-sign vortices, we have two solutions: a cyclotron branch with rotational frequency $\Omega_+$ and a massless branch with $\Omega_-$.
The cyclotron branch $\Omega_+$ is a clear distinction from the result of the massless PVM but is typically unstable with a large Mach number beyond the incompressible approximation in bosonic superfluids (Fig.~\ref{fig:angular frequency}), 
whereas it can be stabilized in fermionic superfluids.
For the massless branch $\Omega_-$, 
it is revealed that,
by ignoring the radial oscillation with small amplitude,
massive vortices move faster or slower than massless vortices along the circular orbit, 
depending on the initial velocity along the circular trajectory.
In other words, the impact of the vortex mass on the trajectory appears as an angle shift, which enables us to quantitatively verify the vortex mass in the experiments (Fig.~\ref{fig:angle_r0_10}).
The intervortex distance of the circular motion has a lower limit $2\sigma$, 
which characterizes the final state of the splitting instability of a doubly-quantized vortex at zero temperature.
We also showed that a vortex--antivortex pair spontaneously annihilates even at absolute zero if the relative kinetic energy is given to vortices with distance $r\leq\sigma$ of translational motion at the initial state.

The impact of the vortex mass is detectable, for example, by tracking the angles of the circular trajectories of like-sign vortices with opposite initial velocities in superfluid atomic gases in a box potential with slab geometry.
The initial velocity can be manipulated by introducing two blue-detuned pinning potentials~\cite{kwon2021sound,samson2016deterministic} that move along a circle with diameter $r>2\sigma$.
Revisiting the problem of splitting instability is another interesting direction for examining the vortex mass.
Therefore, it is helpful to observe the final distance $r_{\mathrm{fin}}$ in both bosonic and fermionic superfluids for evaluating the upper bound of the vortex mass quantitatively with Eq.~\eqref{eq:upper limit of vortex mass}.

Because broken U(1)-symmetry of a superfluid in the ground state is essential for describing the incompressible potential flows within the massive PVM, our theory is also applicable to certain types of multicomponent superfluids.
In addition to segregated binary BECs proposed in Refs.~\cite{richaud2020vortices,richaud2021dynamics,ruban2022direct,matteo2023massive,richaud2023massive,bellettini2024rotational,d2024stability,richaud2024suppression,richaud2024dynamical}, a strongly ferromagnetic superfluid of $^7$Li atoms with a negative quadratic Zeeman energy~\cite{huh2024universality} is a timely target of our theory, in which a vortex in a spin component is occupied by another spin component, thus forming integer or fractional magnetic skyrmions~\cite{huh2024beyond}.
Intriguingly, the cyclotron branch can be stable with a small Mach number in topological quantum fluid $^3$He-B, 
where the vortex core is occupied by superfluid orders other than the bulk order~\cite{lounasmaa1999vortices,vollhardt2013superfluid}; 
thus, the vortex mass can be large.

\begin{acknowledgments}
We thank L. Levrouw, J. Tempere, T. Simula, Y. Shin, and G. Roati for useful discussions. 
H.T is supported by JSPS KAKENHI Grants No. JP18KK0391, and No. JP20H01842; and JST, PRESTO (Japan) Grant No. JPMJPR23O5.
\end{acknowledgments}

\appendix
\section{\label{app:finite size effect}Finite size effect for a vortex--vortex pair}
Here, we show the Hamiltonian of the relative motion of a vortex--vortex pair with a round boundary (see Ref.~\cite{richaud2020vortices}).
We consider a circular boundary with radius $R_{\mathrm{sys}}$ and select its center as the origin of the coordinate plane.
When the boundary is at rest, the normal component of the velocity field with respect to the boundary must be zero at the boundary for the ideal fluid~\cite{batchelor1967introduction}.
According to the method of images in electromagnetics~\cite{jackson2021classical}, an image vortex, which is the counterpart of an image charge, is required to replicate the boundary condition.
In our case, the image vortex of the $i$th ($i=1,2$) vortex at $\bm{r}_i$ with $+\kappa$ must be placed at 
\begin{equation}
    \bm{r}'_i=\frac{R_{\mathrm{sys}}^2}{|\bm{r}_i|^2}\bm{r}_i
\end{equation}
and have opposite circulations $-\kappa$.
Then, one vortex interacts with the three vortices; the other vortex and two image vortices.
The equation of motion of the $i$th vortex is 
\begin{equation}
    m_v\frac{d^2\bm{r}_i}{dt^2}
    =\rho\kappa\bm{e}_z\times\left(\frac{d\bm{r}_i}{dt}-\bm{v}'_{ij}\right) \quad (j\neq i)
\end{equation}
with induced velocity
\begin{align}
    \bm{v}'_{ij}
    &=\frac{\kappa}{2\pi}\frac{\bm{e}_z\times(\bm{r}_i-\bm{r}_j)}{|\bm{r}_i-\bm{r}_j|^2} \notag \\
    &\quad -\frac{\kappa}{2\pi}\frac{\bm{e}_z\times(\bm{r}_i-\bm{r}'_i)}{|\bm{r}_i-\bm{r}'_i|^2}
     -\frac{\kappa}{2\pi}\frac{\bm{e}_z\times(\bm{r}_i-\bm{r}'_j)}{|\bm{r}_i-\bm{r}'_j|^2}.
\end{align}
Subsequently, we introduce the following Lagrangian: 
\begin{align}
    &L'(\bm{r}_1,\bm{r}_2,\dot{\bm{r}}_1,\dot{\bm{r}}_2) \notag \\
    &=\frac{1}{2}m_v(\dot{\bm{r}}^2_1+\dot{\bm{r}}^2_2)
      +\frac{\rho\kappa}{2}(\dot{\bm{r}}_1\times\bm{r}_1+\dot{\bm{r}}_2\times\bm{r}_2)\cdot\bm{e}_z \notag \\
    &\quad+\frac{\rho\kappa^2}{2\pi}\ln\frac{|\bm{r}_1-\bm{r}_2|}{\sigma}
               -\frac{\rho\kappa^2}{2\pi}\ln\left(1-\frac{\bm{r}_1\cdot\bm{r}_2}{R^2_{\mathrm{sys}}}\right) \notag \\
    &\quad\quad-\frac{\rho\kappa^2}{4\pi}\ln\left(1-\frac{|\bm{r}_1|^2}{R^2_{\mathrm{sys}}}\right)
                    -\frac{\rho\kappa^2}{4\pi}\ln\left(1-\frac{|\bm{r}_2|^2}{R^2_{\mathrm{sys}}}\right),
    \label{eq:L'}
\end{align}
where the constants are added and included in each logarithmic term.

Similar to the calculations presented in Sec.~\ref{sec:effective radial potential}, we introduce $\bm{r}_G=(\bm{r}_1+\bm{r}_2)/2$ and $\bm{r}_R=\bm{r}_2-\bm{r}_1$.
We set $\bm{r}_G(t)=0$ by providing the proper initial condition based on the twofold rotational symmetry of the system with respect to the origin.
Therefore, the Lagrangian of relative motion is obtained by substituting $\bm{r}_1=-\bm{r}_R/2$ and $\bm{r}_2=+\bm{r}_R/2$ into Eq.~\eqref{eq:L'}: 
\begin{align}
    &L'_R(\bm{r}_R,\dot{\bm{r}}_R) \notag \\ 
    &=L_R(\bm{r}_R,\dot{\bm{r}}_R)
     -\frac{\rho\kappa^2}{2\pi}\ln\left[1+\frac{|\bm{r}_R|^2}{(2R_{\mathrm{sys}})^2}\right] \notag \\
    &\quad-\frac{\rho\kappa^2}{2\pi}\ln\left[1-\frac{|\bm{r}_R|^2}{(2R_{\mathrm{sys}})^2}\right] \notag \\
    &=L_R(\bm{r}_R,\dot{\bm{r}}_R)
     -\frac{\rho\kappa^2}{2\pi}\ln\left[1-\frac{|\bm{r}_R|^4}{(2R_{\mathrm{sys}})^4}\right],
\end{align}
where $L_R$ denotes the Lagrangian without a boundary, given by Eq.~\eqref{eq:L_R}.
Consequently, the factor caused by the finite-size effect is summarized in an additional logarithmic term.
Using the polar coordinate $\bm{r}_R=(r,\theta)$, we obtain the Lagrangian 
\begin{align}
    &L'_R(r,\theta,\dot{r},\dot{\theta}) \notag \\
    &=L_R(r,\theta,\dot{r},\dot{\theta})
    -\frac{\rho\kappa^2}{2\pi}\ln\left[1-\left(\frac{r}{2R_{\mathrm{sys}}}\right)^4\right]
\end{align}
and the canonical momenta
\begin{align}
    p'_r
    &=\frac{\partial L'_R}{\partial\dot{r}}
     =\frac{\partial L_R}{\partial\dot{r}}
     =p_r, \\
    p'_\theta
    &=\frac{\partial L'_R}{\partial\dot{\theta}}
     =\frac{\partial L_R}{\partial\dot{\theta}}
     =p_\theta
\end{align}
with Eqs.~\eqref{eq:radial momentum} and \eqref{eq:azimuth momentum}.
Therefore, by performing the Legendre transformation $H'_R=\dot{r}p'_r+\dot{\theta}p'_\theta-L'_R$, we obtain the Hamiltonian given by Eq.~\eqref{eq:H'_R}.

\section{\label{app:canonical transformation}Hamiltonian of a vortex--antivortex pair}
Here, we derive the Hamiltonian of a vortex--antivortex pair, as shown in Eq.~\eqref{eq:Hamiltonian for vortex-antivortex} (see Ref.~\cite{ragazzo1994motion}).
According to Eqs.~\eqref{eq:EOM for vortex-antivortex,G} and \eqref{eq:EOM for vortex-antivortex,R}, we introduce the Lagrangian
\begin{align}
    &L(\bm{r}_G,\bm{r}_R,\dot{\bm{r}}_G,\dot{\bm{r}}_R) \notag \\
    &=\frac{1}{2}M_v\dot{\bm{r}}_G^2+\frac{1}{2}\mu_v\dot{\bm{r}}_R^2
    -\frac{\rho\kappa}{2}(\dot{\bm{r}}_G\times\bm{r}_R+\dot{\bm{r}}_R\times\bm{r}_G)\cdot\bm{e}_z \notag \\
    &\quad-\frac{\rho\kappa^2}{2\pi}\ln\frac{|\bm{r}_R|}{\sigma},
\end{align}
where a constant is added and included in the last term.
The canonical momenta are
\begin{align}
    \bm{p}'_G&=\frac{\partial L}{\partial\dot{\bm{r}}_G}=M_v\dot{\bm{r}}_G-\frac{\rho\kappa}{2}\bm{r}_R\times\bm{e}_z, \\ 
    \bm{p}'_R&=\frac{\partial L}{\partial\dot{\bm{r}}_R}=\mu_v\dot{\bm{r}}_R-\frac{\rho\kappa}{2}\bm{r}_G\times\bm{e}_z.
\end{align}
The Hamiltonian is defined by the Legendre transformation $H'=\dot{\bm{r}}_G\cdot\bm{p}'_G+\dot{\bm{r}}_R\cdot\bm{p}'_R-L$ and we obtain
\begin{align}
    &H'(\bm{r}_G,\bm{r}_R,\bm{p}'_G,\bm{p}'_R) \notag \\
    &=\frac{1}{2M_v}\bm{p}'^2_G+\frac{1}{2\mu_v}\bm{p}'^2_R
    +\frac{\rho^2\kappa^2}{2M_v}\bm{r}_G^2+\frac{\rho^2\kappa^2}{32\mu_v}\bm{r}_R^2 \notag \\ 
    &\quad+\frac{\rho\kappa}{2\mu_v}\bm{p}'_R\cdot(\bm{r}_G\times\bm{e}_z)+\frac{\rho\kappa}{2M_v}\bm{p}_G'\cdot(\bm{r}_R\times\bm{e}_z) \notag \\
    &\quad\quad+\frac{\rho\kappa^2}{2\pi}\ln\frac{|\bm{r}_R|}{\sigma}.
\end{align}

Here, we consider performing a canonical transformation from the old momenta $(\bm{p}'_G,\bm{p}'_R)$ to the new momenta $(\bm{p}_G,\bm{p}_R)$. 
The following generating function is introduced:
\begin{equation}
    W=\bm{r}_G\cdot\bm{p}_G+\bm{r}_R\cdot\bm{p}_R-\frac{\rho\kappa}{2}(\bm{r}_G\times\bm{e}_z)\cdot\bm{r}_R.
\end{equation}
Then, relationships between $(\bm{p}'_G,\bm{p}'_R)$ and $(\bm{p}_G,\bm{p}_R)$ are
\begin{align}
    \bm{p}'_G&=\frac{\partial W}{\partial\bm{r}_G}=\bm{p}_G+\frac{\rho\kappa}{2}\bm{r}_R\times\bm{e}_z, \\
    \bm{p}'_R&=\frac{\partial W}{\partial\bm{r}_R}=\bm{p}_R-\frac{\rho\kappa}{2}\bm{r}_G\times\bm{e}_z.
\end{align}
Therefore, we have the new Hamiltonian 
\begin{align}
    &H(\bm{r}_G,\bm{r}_R,\bm{p}_G,\bm{p}_R) \notag \\
    &=H'(\bm{r}_G,\bm{r}_R,\bm{p}'_G(\bm{r}_R,\bm{p}_G),\bm{p}'_R(\bm{r}_G,\bm{p}_R))+\frac{\partial W}{\partial t} \notag \\ 
    &=\frac{\bm{p}_R^2}{2\mu_v}
       +\frac{\rho^2\kappa^2}{8\mu_v}\left(\bm{r}_R+\bm{e}_z\times\frac{\bm{p}_G}{\rho\kappa}\right)^2
       +\frac{\rho\kappa^2}{2\pi}\ln\frac{|\bm{r}_R|}{\sigma},
       \label{eq:Hamiltonian for vortex-antivortex, appendix}
\end{align}
which correspond to Eqs.~\eqref{eq:Hamiltonian for vortex-antivortex} and \eqref{eq:effective relative potential}.
For reference, we show Hamilton's equations: 
\begin{align}
    \dot{\bm{r}}_G
    &=\frac{\partial H}{\partial \bm{p}_G}
     =\frac{\partial U_{\mathrm{eff}}}{\partial \bm{p}_G}, \\
    \dot{\bm{r}}_R
    &=\frac{\partial H}{\partial \bm{p}_R}
     =\frac{\bm{p}_R}{\mu_v}, \\
    \dot{\bm{p}}_G
    &=-\frac{\partial H}{\partial \bm{r}_G}
     =0,\\
    \dot{\bm{p}}_R
    &=-\frac{\partial H}{\partial \bm{r}_R}
     =-\frac{\partial U_{\mathrm{eff}}}{\partial \bm{r}_R}.
\end{align}
The third equation indicates that the pseudomomentum $\bm{p}_G$ is conserved.

\bibliography{cite}

\begin{thebibliography}{53}%
\makeatletter
\providecommand \@ifxundefined [1]{%
 \@ifx{#1\undefined}
}%
\providecommand \@ifnum [1]{%
 \ifnum #1\expandafter \@firstoftwo
 \else \expandafter \@secondoftwo
 \fi
}%
\providecommand \@ifx [1]{%
 \ifx #1\expandafter \@firstoftwo
 \else \expandafter \@secondoftwo
 \fi
}%
\providecommand \natexlab [1]{#1}%
\providecommand \enquote  [1]{``#1''}%
\providecommand \bibnamefont  [1]{#1}%
\providecommand \bibfnamefont [1]{#1}%
\providecommand \citenamefont [1]{#1}%
\providecommand \href@noop [0]{\@secondoftwo}%
\providecommand \href [0]{\begingroup \@sanitize@url \@href}%
\providecommand \@href[1]{\@@startlink{#1}\@@href}%
\providecommand \@@href[1]{\endgroup#1\@@endlink}%
\providecommand \@sanitize@url [0]{\catcode `\\12\catcode `\$12\catcode `\&12\catcode `\#12\catcode `\^12\catcode `\_12\catcode `\%12\relax}%
\providecommand \@@startlink[1]{}%
\providecommand \@@endlink[0]{}%
\providecommand \url  [0]{\begingroup\@sanitize@url \@url }%
\providecommand \@url [1]{\endgroup\@href {#1}{\urlprefix }}%
\providecommand \urlprefix  [0]{URL }%
\providecommand \Eprint [0]{\href }%
\providecommand \doibase [0]{https://doi.org/}%
\providecommand \selectlanguage [0]{\@gobble}%
\providecommand \bibinfo  [0]{\@secondoftwo}%
\providecommand \bibfield  [0]{\@secondoftwo}%
\providecommand \translation [1]{[#1]}%
\providecommand \BibitemOpen [0]{}%
\providecommand \bibitemStop [0]{}%
\providecommand \bibitemNoStop [0]{.\EOS\space}%
\providecommand \EOS [0]{\spacefactor3000\relax}%
\providecommand \BibitemShut  [1]{\csname bibitem#1\endcsname}%
\let\auto@bib@innerbib\@empty
%</preamble>
\bibitem [{\citenamefont {Batchelor}(1967)}]{batchelor1967introduction}%
  \BibitemOpen
  \bibfield  {author} {\bibinfo {author} {\bibfnamefont {G.~K.}\ \bibnamefont {Batchelor}},\ }\href@noop {} {\emph {\bibinfo {title} {{An Introduction to Fluid Dynamics}}}}\ (\bibinfo  {publisher} {Cambridge university press, Cambridge},\ \bibinfo {year} {1967})\BibitemShut {NoStop}%
\bibitem [{\citenamefont {Donnelly}(1991)}]{donnelly1991quantized}%
  \BibitemOpen
  \bibfield  {author} {\bibinfo {author} {\bibfnamefont {R.~J.}\ \bibnamefont {Donnelly}},\ }\href@noop {} {\emph {\bibinfo {title} {{Quantized Vortices in Helium II}}}},\ Vol.~\bibinfo {volume} {2}\ (\bibinfo  {publisher} {Cambridge University Press, Cambridge},\ \bibinfo {year} {1991})\BibitemShut {NoStop}%
\bibitem [{\citenamefont {Freilich}\ \emph {et~al.}(2010)\citenamefont {Freilich}, \citenamefont {Bianchi}, \citenamefont {Kaufman}, \citenamefont {Langin},\ and\ \citenamefont {Hall}}]{freilich2010real}%
  \BibitemOpen
  \bibfield  {author} {\bibinfo {author} {\bibfnamefont {D.~V.}\ \bibnamefont {Freilich}}, \bibinfo {author} {\bibfnamefont {D.~M.}\ \bibnamefont {Bianchi}}, \bibinfo {author} {\bibfnamefont {A.~M.}\ \bibnamefont {Kaufman}}, \bibinfo {author} {\bibfnamefont {T.~K.}\ \bibnamefont {Langin}},\ and\ \bibinfo {author} {\bibfnamefont {D.~S.}\ \bibnamefont {Hall}},\ }\bibfield  {title} {\bibinfo {title} {{Real-time dynamics of single vortex lines and vortex dipoles in a Bose-Einstein condensate}},\ }\href {https://doi.org/10.1126/science.1191224} {\bibfield  {journal} {\bibinfo  {journal} {Science}\ }\textbf {\bibinfo {volume} {329}},\ \bibinfo {pages} {1182} (\bibinfo {year} {2010})}\BibitemShut {NoStop}%
\bibitem [{\citenamefont {Kwon}\ \emph {et~al.}(2014)\citenamefont {Kwon}, \citenamefont {Moon}, \citenamefont {Choi}, \citenamefont {Seo},\ and\ \citenamefont {Shin}}]{kwon2014relaxation}%
  \BibitemOpen
  \bibfield  {author} {\bibinfo {author} {\bibfnamefont {W.~J.}\ \bibnamefont {Kwon}}, \bibinfo {author} {\bibfnamefont {G.}~\bibnamefont {Moon}}, \bibinfo {author} {\bibfnamefont {J.-y.}\ \bibnamefont {Choi}}, \bibinfo {author} {\bibfnamefont {S.~W.}\ \bibnamefont {Seo}},\ and\ \bibinfo {author} {\bibfnamefont {Y.-i.}\ \bibnamefont {Shin}},\ }\bibfield  {title} {\bibinfo {title} {{Relaxation of superfluid turbulence in highly oblate Bose-Einstein condensates}},\ }\href {https://doi.org/10.1103/PhysRevA.90.063627} {\bibfield  {journal} {\bibinfo  {journal} {Phys. Rev. A}\ }\textbf {\bibinfo {volume} {90}},\ \bibinfo {pages} {063627} (\bibinfo {year} {2014})}\BibitemShut {NoStop}%
\bibitem [{\citenamefont {Moon}\ \emph {et~al.}(2015)\citenamefont {Moon}, \citenamefont {Kwon}, \citenamefont {Lee},\ and\ \citenamefont {Shin}}]{moon2015thermal}%
  \BibitemOpen
  \bibfield  {author} {\bibinfo {author} {\bibfnamefont {G.}~\bibnamefont {Moon}}, \bibinfo {author} {\bibfnamefont {W.~J.}\ \bibnamefont {Kwon}}, \bibinfo {author} {\bibfnamefont {H.}~\bibnamefont {Lee}},\ and\ \bibinfo {author} {\bibfnamefont {Y.-i.}\ \bibnamefont {Shin}},\ }\bibfield  {title} {\bibinfo {title} {{Thermal friction on quantum vortices in a Bose-Einstein condensate}},\ }\href {https://doi.org/10.1103/PhysRevA.92.051601} {\bibfield  {journal} {\bibinfo  {journal} {Phys. Rev. A}\ }\textbf {\bibinfo {volume} {92}},\ \bibinfo {pages} {051601} (\bibinfo {year} {2015})}\BibitemShut {NoStop}%
\bibitem [{\citenamefont {Kwon}\ \emph {et~al.}(2016)\citenamefont {Kwon}, \citenamefont {Kim}, \citenamefont {Seo},\ and\ \citenamefont {Shin}}]{kwon2016observation}%
  \BibitemOpen
  \bibfield  {author} {\bibinfo {author} {\bibfnamefont {W.~J.}\ \bibnamefont {Kwon}}, \bibinfo {author} {\bibfnamefont {J.~H.}\ \bibnamefont {Kim}}, \bibinfo {author} {\bibfnamefont {S.~W.}\ \bibnamefont {Seo}},\ and\ \bibinfo {author} {\bibfnamefont {Y.}~\bibnamefont {Shin}},\ }\bibfield  {title} {\bibinfo {title} {{Observation of von K\'arm\'an Vortex Street in an Atomic Superfluid Gas}},\ }\href {https://doi.org/10.1103/PhysRevLett.117.245301} {\bibfield  {journal} {\bibinfo  {journal} {Phys. Rev. Lett.}\ }\textbf {\bibinfo {volume} {117}},\ \bibinfo {pages} {245301} (\bibinfo {year} {2016})}\BibitemShut {NoStop}%
\bibitem [{\citenamefont {Saffman}(1995)}]{saffman1995vortex}%
  \BibitemOpen
  \bibfield  {author} {\bibinfo {author} {\bibfnamefont {P.~G.}\ \bibnamefont {Saffman}},\ }\href@noop {} {\emph {\bibinfo {title} {{Vortex Dynamics}}}}\ (\bibinfo  {publisher} {Cambridge university press, Cambridge},\ \bibinfo {year} {1995})\BibitemShut {NoStop}%
\bibitem [{\citenamefont {Ragazzo}\ \emph {et~al.}(1994)\citenamefont {Ragazzo}, \citenamefont {Koiller},\ and\ \citenamefont {Oliva}}]{ragazzo1994motion}%
  \BibitemOpen
  \bibfield  {author} {\bibinfo {author} {\bibfnamefont {C.~G.}\ \bibnamefont {Ragazzo}}, \bibinfo {author} {\bibfnamefont {J.}~\bibnamefont {Koiller}},\ and\ \bibinfo {author} {\bibfnamefont {W.}~\bibnamefont {Oliva}},\ }\bibfield  {title} {\bibinfo {title} {{On the motion of two-dimensional vortices with mass}},\ }\href {https://doi.org/10.1007/BF02430639} {\bibfield  {journal} {\bibinfo  {journal} {J. Nonlinear Sci.}\ }\textbf {\bibinfo {volume} {4}},\ \bibinfo {pages} {375} (\bibinfo {year} {1994})}\BibitemShut {NoStop}%
\bibitem [{\citenamefont {Fischer}(1999)}]{fischer1999motion}%
  \BibitemOpen
  \bibfield  {author} {\bibinfo {author} {\bibfnamefont {U.~R.}\ \bibnamefont {Fischer}},\ }\bibfield  {title} {\bibinfo {title} {{Motion of quantized vortices as elementary objects}},\ }\href {https://doi.org/https://doi.org/10.1006/aphy.1999.5969} {\bibfield  {journal} {\bibinfo  {journal} {Ann. Phys. (NY)}\ }\textbf {\bibinfo {volume} {278}},\ \bibinfo {pages} {62} (\bibinfo {year} {1999})}\BibitemShut {NoStop}%
\bibitem [{\citenamefont {Kopnin}(1978)}]{kopnin1978frequency}%
  \BibitemOpen
  \bibfield  {author} {\bibinfo {author} {\bibfnamefont {N.}~\bibnamefont {Kopnin}},\ }\bibfield  {title} {\bibinfo {title} {{Frequency singularities of the dissipation in the mixed state of pure type-II superconductors at low temperatures}},\ }\bibfield  {journal} {\bibinfo  {journal} {JETP Lett.(USSR)(Engl. Transl.);(United States)}\ }\textbf {\bibinfo {volume} {27}},\ \href {https://www.osti.gov/biblio/6918051} {} (\bibinfo {year} {1978})\BibitemShut {NoStop}%
\bibitem [{\citenamefont {Kopnin}\ and\ \citenamefont {Vinokur}(1998)}]{kopnin1998dynamic}%
  \BibitemOpen
  \bibfield  {author} {\bibinfo {author} {\bibfnamefont {N.~B.}\ \bibnamefont {Kopnin}}\ and\ \bibinfo {author} {\bibfnamefont {V.~M.}\ \bibnamefont {Vinokur}},\ }\bibfield  {title} {\bibinfo {title} {{Dynamic vortex mass in clean fermi superfluids and superconductors}},\ }\href {https://doi.org/10.1103/PhysRevLett.81.3952} {\bibfield  {journal} {\bibinfo  {journal} {Phys. Rev. Lett.}\ }\textbf {\bibinfo {volume} {81}},\ \bibinfo {pages} {3952} (\bibinfo {year} {1998})}\BibitemShut {NoStop}%
\bibitem [{\citenamefont {Simula}(2018)}]{simula2018vortex}%
  \BibitemOpen
  \bibfield  {author} {\bibinfo {author} {\bibfnamefont {T.}~\bibnamefont {Simula}},\ }\bibfield  {title} {\bibinfo {title} {{Vortex mass in a superfluid}},\ }\href {https://doi.org/10.1103/PhysRevA.97.023609} {\bibfield  {journal} {\bibinfo  {journal} {Phys. Rev. A}\ }\textbf {\bibinfo {volume} {97}},\ \bibinfo {pages} {023609} (\bibinfo {year} {2018})}\BibitemShut {NoStop}%
\bibitem [{\citenamefont {Richaud}\ \emph {et~al.}(2020)\citenamefont {Richaud}, \citenamefont {Penna}, \citenamefont {Mayol},\ and\ \citenamefont {Guilleumas}}]{richaud2020vortices}%
  \BibitemOpen
  \bibfield  {author} {\bibinfo {author} {\bibfnamefont {A.}~\bibnamefont {Richaud}}, \bibinfo {author} {\bibfnamefont {V.}~\bibnamefont {Penna}}, \bibinfo {author} {\bibfnamefont {R.}~\bibnamefont {Mayol}},\ and\ \bibinfo {author} {\bibfnamefont {M.}~\bibnamefont {Guilleumas}},\ }\bibfield  {title} {\bibinfo {title} {{Vortices with massive cores in a binary mixture of Bose-Einstein condensates}},\ }\href {https://doi.org/10.1103/PhysRevA.101.013630} {\bibfield  {journal} {\bibinfo  {journal} {Phys. Rev. A}\ }\textbf {\bibinfo {volume} {101}},\ \bibinfo {pages} {013630} (\bibinfo {year} {2020})}\BibitemShut {NoStop}%
\bibitem [{\citenamefont {Richaud}\ \emph {et~al.}(2021)\citenamefont {Richaud}, \citenamefont {Penna},\ and\ \citenamefont {Fetter}}]{richaud2021dynamics}%
  \BibitemOpen
  \bibfield  {author} {\bibinfo {author} {\bibfnamefont {A.}~\bibnamefont {Richaud}}, \bibinfo {author} {\bibfnamefont {V.}~\bibnamefont {Penna}},\ and\ \bibinfo {author} {\bibfnamefont {A.~L.}\ \bibnamefont {Fetter}},\ }\bibfield  {title} {\bibinfo {title} {{Dynamics of massive point vortices in a binary mixture of Bose-Einstein condensates}},\ }\href {https://doi.org/10.1103/PhysRevA.103.023311} {\bibfield  {journal} {\bibinfo  {journal} {Phys. Rev. A}\ }\textbf {\bibinfo {volume} {103}},\ \bibinfo {pages} {023311} (\bibinfo {year} {2021})}\BibitemShut {NoStop}%
\bibitem [{\citenamefont {Ruban}(2022)}]{ruban2022direct}%
  \BibitemOpen
  \bibfield  {author} {\bibinfo {author} {\bibfnamefont {V.~P.}\ \bibnamefont {Ruban}},\ }\bibfield  {title} {\bibinfo {title} {{Direct and reverse precession of a massive vortex in a binary Bose-Einstein condensate}},\ }\href {https://doi.org/10.1134/S0021364022100290} {\bibfield  {journal} {\bibinfo  {journal} {JETP Letters}\ }\textbf {\bibinfo {volume} {115}},\ \bibinfo {pages} {415} (\bibinfo {year} {2022})}\BibitemShut {NoStop}%
\bibitem [{\citenamefont {Caldara}\ \emph {et~al.}(2023)\citenamefont {Caldara}, \citenamefont {Richaud}, \citenamefont {Capone},\ and\ \citenamefont {Massignan}}]{matteo2023massive}%
  \BibitemOpen
  \bibfield  {author} {\bibinfo {author} {\bibfnamefont {M.}~\bibnamefont {Caldara}}, \bibinfo {author} {\bibfnamefont {A.}~\bibnamefont {Richaud}}, \bibinfo {author} {\bibfnamefont {M.}~\bibnamefont {Capone}},\ and\ \bibinfo {author} {\bibfnamefont {P.}~\bibnamefont {Massignan}},\ }\bibfield  {title} {\bibinfo {title} {{Massive superfluid vortices and vortex necklaces on a planar annulus}},\ }\href {https://doi.org/10.21468/SciPostPhys.15.2.057} {\bibfield  {journal} {\bibinfo  {journal} {SciPost Phys.}\ }\textbf {\bibinfo {volume} {15}},\ \bibinfo {pages} {057} (\bibinfo {year} {2023})}\BibitemShut {NoStop}%
\bibitem [{\citenamefont {Richaud}\ \emph {et~al.}(2023{\natexlab{a}})\citenamefont {Richaud}, \citenamefont {Penna},\ and\ \citenamefont {Fetter}}]{richaud2023massive}%
  \BibitemOpen
  \bibfield  {author} {\bibinfo {author} {\bibfnamefont {A.}~\bibnamefont {Richaud}}, \bibinfo {author} {\bibfnamefont {V.}~\bibnamefont {Penna}},\ and\ \bibinfo {author} {\bibfnamefont {A.~L.}\ \bibnamefont {Fetter}},\ }\bibfield  {title} {\bibinfo {title} {{Massive quantum vortices in superfluids}},\ }\href {https://doi.org/10.1088/1742-6596/2494/1/012016} {\bibfield  {journal} {\bibinfo  {journal} {J. Phys.: Conf. Ser.}\ }\textbf {\bibinfo {volume} {2494}},\ \bibinfo {pages} {012016} (\bibinfo {year} {2023}{\natexlab{a}})}\BibitemShut {NoStop}%
\bibitem [{\citenamefont {Bellettini}\ \emph {et~al.}(2024)\citenamefont {Bellettini}, \citenamefont {Richaud},\ and\ \citenamefont {Penna}}]{bellettini2024rotational}%
  \BibitemOpen
  \bibfield  {author} {\bibinfo {author} {\bibfnamefont {A.}~\bibnamefont {Bellettini}}, \bibinfo {author} {\bibfnamefont {A.}~\bibnamefont {Richaud}},\ and\ \bibinfo {author} {\bibfnamefont {V.}~\bibnamefont {Penna}},\ }\bibfield  {title} {\bibinfo {title} {{Rotational states of an asymmetric vortex pair with mass imbalance in binary condensates}},\ }\href {https://doi.org/10.1103/PhysRevA.109.053301} {\bibfield  {journal} {\bibinfo  {journal} {Phys. Rev. A}\ }\textbf {\bibinfo {volume} {109}},\ \bibinfo {pages} {053301} (\bibinfo {year} {2024})}\BibitemShut {NoStop}%
\bibitem [{\citenamefont {D'Ambroise}\ \emph {et~al.}()\citenamefont {D'Ambroise}, \citenamefont {Wang}, \citenamefont {Ticknor}, \citenamefont {Carretero-Gonz{\'a}lez},\ and\ \citenamefont {Kevrekidis}}]{d2024stability}%
  \BibitemOpen
  \bibfield  {author} {\bibinfo {author} {\bibfnamefont {J.}~\bibnamefont {D'Ambroise}}, \bibinfo {author} {\bibfnamefont {W.}~\bibnamefont {Wang}}, \bibinfo {author} {\bibfnamefont {C.}~\bibnamefont {Ticknor}}, \bibinfo {author} {\bibfnamefont {R.}~\bibnamefont {Carretero-Gonz{\'a}lez}},\ and\ \bibinfo {author} {\bibfnamefont {P.}~\bibnamefont {Kevrekidis}},\ }\bibfield  {title} {\bibinfo {title} {{Stability and dynamics of massive vortices in two-component Bose-Einstein condensates}},\ }\href {https://doi.org/10.48550/arXiv.2407.10324} {\bibinfo  {journal} {arXiv:2407.10324}\ }\BibitemShut {NoStop}%
\bibitem [{\citenamefont {Caldara}\ \emph {et~al.}(2024)\citenamefont {Caldara}, \citenamefont {Richaud}, \citenamefont {Capone},\ and\ \citenamefont {Massignan}}]{richaud2024suppression}%
  \BibitemOpen
\bibfield  {journal} {  }\bibfield  {author} {\bibinfo {author} {\bibfnamefont {M.}~\bibnamefont {Caldara}}, \bibinfo {author} {\bibfnamefont {A.}~\bibnamefont {Richaud}}, \bibinfo {author} {\bibfnamefont {M.}~\bibnamefont {Capone}},\ and\ \bibinfo {author} {\bibfnamefont {P.}~\bibnamefont {Massignan}},\ }\bibfield  {title} {\bibinfo {title} {{Suppression of the superfluid Kelvin-Helmholtz instability due to massive vortex cores, friction and confinement}},\ }\href {https://doi.org/10.21468/SciPostPhys.17.3.076} {\bibfield  {journal} {\bibinfo  {journal} {SciPost Phys.}\ }\textbf {\bibinfo {volume} {17}},\ \bibinfo {pages} {076} (\bibinfo {year} {2024})}\BibitemShut {NoStop}%
\bibitem [{\citenamefont {Richaud}\ \emph {et~al.}()\citenamefont {Richaud}, \citenamefont {Caldara}, \citenamefont {Capone}, \citenamefont {Massignan},\ and\ \citenamefont {Wlaz{\l}owski}}]{richaud2024dynamical}%
  \BibitemOpen
  \bibfield  {author} {\bibinfo {author} {\bibfnamefont {A.}~\bibnamefont {Richaud}}, \bibinfo {author} {\bibfnamefont {M.}~\bibnamefont {Caldara}}, \bibinfo {author} {\bibfnamefont {M.}~\bibnamefont {Capone}}, \bibinfo {author} {\bibfnamefont {P.}~\bibnamefont {Massignan}},\ and\ \bibinfo {author} {\bibfnamefont {G.}~\bibnamefont {Wlaz{\l}owski}},\ }\bibfield  {title} {\bibinfo {title} {{Dynamical signature of vortex mass in Fermi superfluids}},\ }\href {https://doi.org/10.48550/arXiv.2410.12417} {\bibinfo  {journal} {arXiv:2410.12417}\ }\BibitemShut {NoStop}%
\bibitem [{\citenamefont {Baym}\ and\ \citenamefont {Chandler}(1983)}]{baym1983hydrodynamics}%
  \BibitemOpen
\bibfield  {journal} {  }\bibfield  {author} {\bibinfo {author} {\bibfnamefont {G.}~\bibnamefont {Baym}}\ and\ \bibinfo {author} {\bibfnamefont {E.}~\bibnamefont {Chandler}},\ }\bibfield  {title} {\bibinfo {title} {{The hydrodynamics of rotating superfluids. I. Zero-temperature, nondissipative theory}},\ }\href {https://doi.org/10.1007/BF00681839} {\bibfield  {journal} {\bibinfo  {journal} {J. Low Temp. Phys.}\ }\textbf {\bibinfo {volume} {50}},\ \bibinfo {pages} {57} (\bibinfo {year} {1983})}\BibitemShut {NoStop}%
\bibitem [{\citenamefont {Popov}(1973)}]{popov1973quantum}%
  \BibitemOpen
  \bibfield  {author} {\bibinfo {author} {\bibfnamefont {V.}~\bibnamefont {Popov}},\ }\bibfield  {title} {\bibinfo {title} {{Quantum vortices and phase transitions in Bose systems}},\ }\href@noop {} {\bibfield  {journal} {\bibinfo  {journal} {Zh. Eksp. Teor. Fiz.}\ }\textbf {\bibinfo {volume} {64}},\ \bibinfo {pages} {672} (\bibinfo {year} {1973})},\ \bibinfo {note} {[Sov. Phys. JETP \textbf{37}, 341 (1973)]}\BibitemShut {NoStop}%
\bibitem [{\citenamefont {Duan}\ and\ \citenamefont {Leggett}(1992)}]{duan1992inertial}%
  \BibitemOpen
  \bibfield  {author} {\bibinfo {author} {\bibfnamefont {J.-M.}\ \bibnamefont {Duan}}\ and\ \bibinfo {author} {\bibfnamefont {A.~J.}\ \bibnamefont {Leggett}},\ }\bibfield  {title} {\bibinfo {title} {{Inertial mass of a moving singularity in a Fermi superfluid}},\ }\href {https://doi.org/10.1103/PhysRevLett.68.1216} {\bibfield  {journal} {\bibinfo  {journal} {Phys. Rev. Lett.}\ }\textbf {\bibinfo {volume} {68}},\ \bibinfo {pages} {1216} (\bibinfo {year} {1992})}\BibitemShut {NoStop}%
\bibitem [{\citenamefont {Duan}(1994)}]{duan1994mass}%
  \BibitemOpen
  \bibfield  {author} {\bibinfo {author} {\bibfnamefont {J.-M.}\ \bibnamefont {Duan}},\ }\bibfield  {title} {\bibinfo {title} {{Mass of a vortex line in superfluid $^{4}\mathrm{He}$: Effects of gauge-symmetry breaking}},\ }\href {https://doi.org/10.1103/PhysRevB.49.12381} {\bibfield  {journal} {\bibinfo  {journal} {Phys. Rev. B}\ }\textbf {\bibinfo {volume} {49}},\ \bibinfo {pages} {12381} (\bibinfo {year} {1994})}\BibitemShut {NoStop}%
\bibitem [{\citenamefont {Volovik}(2003)}]{volovik2003universe}%
  \BibitemOpen
  \bibfield  {author} {\bibinfo {author} {\bibfnamefont {G.~E.}\ \bibnamefont {Volovik}},\ }\href@noop {} {\emph {\bibinfo {title} {{The Universe in a Helium Droplet}}}},\ Vol.\ \bibinfo {volume} {117}\ (\bibinfo  {publisher} {Oxford University Press, Oxford},\ \bibinfo {year} {2003})\BibitemShut {NoStop}%
\bibitem [{\citenamefont {Kwon}\ \emph {et~al.}(2021)\citenamefont {Kwon}, \citenamefont {Del~Pace}, \citenamefont {Xhani}, \citenamefont {Galantucci}, \citenamefont {Muzi~Falconi}, \citenamefont {Inguscio}, \citenamefont {Scazza},\ and\ \citenamefont {Roati}}]{kwon2021sound}%
  \BibitemOpen
  \bibfield  {author} {\bibinfo {author} {\bibfnamefont {W.~J.}\ \bibnamefont {Kwon}}, \bibinfo {author} {\bibfnamefont {G.}~\bibnamefont {Del~Pace}}, \bibinfo {author} {\bibfnamefont {K.}~\bibnamefont {Xhani}}, \bibinfo {author} {\bibfnamefont {L.}~\bibnamefont {Galantucci}}, \bibinfo {author} {\bibfnamefont {A.}~\bibnamefont {Muzi~Falconi}}, \bibinfo {author} {\bibfnamefont {M.}~\bibnamefont {Inguscio}}, \bibinfo {author} {\bibfnamefont {F.}~\bibnamefont {Scazza}},\ and\ \bibinfo {author} {\bibfnamefont {G.}~\bibnamefont {Roati}},\ }\bibfield  {title} {\bibinfo {title} {{Sound emission and annihilations in a programmable quantum vortex collider}},\ }\href {https://doi.org/10.1038/s41586-021-04047-4} {\bibfield  {journal} {\bibinfo  {journal} {Nature(London)}\ }\textbf {\bibinfo {volume} {600}},\ \bibinfo {pages} {64} (\bibinfo {year} {2021})}\BibitemShut {NoStop}%
\bibitem [{\citenamefont {Hern{\'a}ndez-Rajkov}\ \emph {et~al.}(2024)\citenamefont {Hern{\'a}ndez-Rajkov}, \citenamefont {Grani}, \citenamefont {Scazza}, \citenamefont {Del~Pace}, \citenamefont {Kwon}, \citenamefont {Inguscio}, \citenamefont {Xhani}, \citenamefont {Fort}, \citenamefont {Modugno}, \citenamefont {Marino} \emph {et~al.}}]{hernandez2024connecting}%
  \BibitemOpen
  \bibfield  {author} {\bibinfo {author} {\bibfnamefont {D.}~\bibnamefont {Hern{\'a}ndez-Rajkov}}, \bibinfo {author} {\bibfnamefont {N.}~\bibnamefont {Grani}}, \bibinfo {author} {\bibfnamefont {F.}~\bibnamefont {Scazza}}, \bibinfo {author} {\bibfnamefont {G.}~\bibnamefont {Del~Pace}}, \bibinfo {author} {\bibfnamefont {W.}~\bibnamefont {Kwon}}, \bibinfo {author} {\bibfnamefont {M.}~\bibnamefont {Inguscio}}, \bibinfo {author} {\bibfnamefont {K.}~\bibnamefont {Xhani}}, \bibinfo {author} {\bibfnamefont {C.}~\bibnamefont {Fort}}, \bibinfo {author} {\bibfnamefont {M.}~\bibnamefont {Modugno}}, \bibinfo {author} {\bibfnamefont {F.}~\bibnamefont {Marino}}, \emph {et~al.},\ }\bibfield  {title} {\bibinfo {title} {{Connecting shear flow and vortex array instabilities in annular atomic superfluids}},\ }\href {https://doi.org/10.1038/s41567-024-02466-4} {\bibfield  {journal} {\bibinfo  {journal} {Nat. Phys.}\ ,\ \bibinfo {pages} {1}} (\bibinfo {year} {2024})}\BibitemShut {NoStop}%
\bibitem [{\citenamefont {Shin}\ \emph {et~al.}(2004)\citenamefont {Shin}, \citenamefont {Saba}, \citenamefont {Vengalattore}, \citenamefont {Pasquini}, \citenamefont {Sanner}, \citenamefont {Leanhardt}, \citenamefont {Prentiss}, \citenamefont {Pritchard},\ and\ \citenamefont {Ketterle}}]{shin2004dynamical}%
  \BibitemOpen
  \bibfield  {author} {\bibinfo {author} {\bibfnamefont {Y.}~\bibnamefont {Shin}}, \bibinfo {author} {\bibfnamefont {M.}~\bibnamefont {Saba}}, \bibinfo {author} {\bibfnamefont {M.}~\bibnamefont {Vengalattore}}, \bibinfo {author} {\bibfnamefont {T.~A.}\ \bibnamefont {Pasquini}}, \bibinfo {author} {\bibfnamefont {C.}~\bibnamefont {Sanner}}, \bibinfo {author} {\bibfnamefont {A.~E.}\ \bibnamefont {Leanhardt}}, \bibinfo {author} {\bibfnamefont {M.}~\bibnamefont {Prentiss}}, \bibinfo {author} {\bibfnamefont {D.~E.}\ \bibnamefont {Pritchard}},\ and\ \bibinfo {author} {\bibfnamefont {W.}~\bibnamefont {Ketterle}},\ }\bibfield  {title} {\bibinfo {title} {{Dynamical instability of a doubly quantized vortex in a Bose-Einstein condensate}},\ }\href {https://doi.org/10.1103/PhysRevLett.93.160406} {\bibfield  {journal} {\bibinfo  {journal} {Phys. Rev. Lett.}\ }\textbf {\bibinfo {volume} {93}},\ \bibinfo {pages} {160406} (\bibinfo {year} {2004})}\BibitemShut {NoStop}%
\bibitem [{\citenamefont {Samson}\ \emph {et~al.}(2016)\citenamefont {Samson}, \citenamefont {Wilson}, \citenamefont {Newman},\ and\ \citenamefont {Anderson}}]{samson2016deterministic}%
  \BibitemOpen
  \bibfield  {author} {\bibinfo {author} {\bibfnamefont {E.~C.}\ \bibnamefont {Samson}}, \bibinfo {author} {\bibfnamefont {K.~E.}\ \bibnamefont {Wilson}}, \bibinfo {author} {\bibfnamefont {Z.~L.}\ \bibnamefont {Newman}},\ and\ \bibinfo {author} {\bibfnamefont {B.~P.}\ \bibnamefont {Anderson}},\ }\bibfield  {title} {\bibinfo {title} {{Deterministic creation, pinning, and manipulation of quantized vortices in a Bose-Einstein condensate}},\ }\href {https://doi.org/10.1103/PhysRevA.93.023603} {\bibfield  {journal} {\bibinfo  {journal} {Phys. Rev. A}\ }\textbf {\bibinfo {volume} {93}},\ \bibinfo {pages} {023603} (\bibinfo {year} {2016})}\BibitemShut {NoStop}%
\bibitem [{\citenamefont {Khalifa}\ and\ \citenamefont {Taha}(2024)}]{khalifa2024vortex}%
  \BibitemOpen
  \bibfield  {author} {\bibinfo {author} {\bibfnamefont {N.~M.}\ \bibnamefont {Khalifa}}\ and\ \bibinfo {author} {\bibfnamefont {H.~E.}\ \bibnamefont {Taha}},\ }\bibfield  {title} {\bibinfo {title} {{Vortex dynamics: A variational approach using the principle of least action}},\ }\href {https://doi.org/10.1103/PhysRevFluids.9.034701} {\bibfield  {journal} {\bibinfo  {journal} {Phys. Rev. Fluids}\ }\textbf {\bibinfo {volume} {9}},\ \bibinfo {pages} {034701} (\bibinfo {year} {2024})}\BibitemShut {NoStop}%
\bibitem [{\citenamefont {Volovik}(1998)}]{volovik1998vortex}%
  \BibitemOpen
  \bibfield  {author} {\bibinfo {author} {\bibfnamefont {G.}~\bibnamefont {Volovik}},\ }\bibfield  {title} {\bibinfo {title} {{Vortex mass in BCS systems: Kopnin and Baym-Chandler contributions}},\ }\href {https://doi.org/10.1134/1.567692} {\bibfield  {journal} {\bibinfo  {journal} {JETP Letters}\ }\textbf {\bibinfo {volume} {67}},\ \bibinfo {pages} {528} (\bibinfo {year} {1998})}\BibitemShut {NoStop}%
\bibitem [{\citenamefont {Pethick}\ and\ \citenamefont {Smith}(2008)}]{pethick2008bose}%
  \BibitemOpen
  \bibfield  {author} {\bibinfo {author} {\bibfnamefont {C.~J.}\ \bibnamefont {Pethick}}\ and\ \bibinfo {author} {\bibfnamefont {H.}~\bibnamefont {Smith}},\ }\href@noop {} {\emph {\bibinfo {title} {{Bose--Einstein Condensation in Dilute Gases}}}}\ (\bibinfo  {publisher} {Cambridge university press, Cambridge},\ \bibinfo {year} {2008})\BibitemShut {NoStop}%
\bibitem [{\citenamefont {Vollhardt}\ and\ \citenamefont {Wolfle}(2013)}]{vollhardt2013superfluid}%
  \BibitemOpen
  \bibfield  {author} {\bibinfo {author} {\bibfnamefont {D.}~\bibnamefont {Vollhardt}}\ and\ \bibinfo {author} {\bibfnamefont {P.}~\bibnamefont {Wolfle}},\ }\href@noop {} {\emph {\bibinfo {title} {{The Superfluid Phases of Helium 3}}}}\ (\bibinfo  {publisher} {Courier Corporation, North Chelmsford},\ \bibinfo {year} {2013})\BibitemShut {NoStop}%
\bibitem [{Note1()}]{Note1}%
  \BibitemOpen
  \bibinfo {note} {Using $\xi \approx 10^3-10^4 \protect \mathrm {\r A}$ and $n\sim 10^{-27} \protect \mathrm {m}^{-3}$ in the superfluid phase of $^3$He-B at zero temperature, we obtain $\protect \tilde {c}_s^F\gg \protect \tilde {c}_s^B$.}\BibitemShut {Stop}%
\bibitem [{\citenamefont {Stockdale}\ \emph {et~al.}(2021)\citenamefont {Stockdale}, \citenamefont {Reeves},\ and\ \citenamefont {Davis}}]{stockdale2021dynamical}%
  \BibitemOpen
  \bibfield  {author} {\bibinfo {author} {\bibfnamefont {O.~R.}\ \bibnamefont {Stockdale}}, \bibinfo {author} {\bibfnamefont {M.~T.}\ \bibnamefont {Reeves}},\ and\ \bibinfo {author} {\bibfnamefont {M.~J.}\ \bibnamefont {Davis}},\ }\bibfield  {title} {\bibinfo {title} {{Dynamical mechanisms of vortex pinning in superfluid thin films}},\ }\href {https://doi.org/10.1103/PhysRevLett.127.255302} {\bibfield  {journal} {\bibinfo  {journal} {Phys. Rev. Lett.}\ }\textbf {\bibinfo {volume} {127}},\ \bibinfo {pages} {255302} (\bibinfo {year} {2021})}\BibitemShut {NoStop}%
\bibitem [{Note2()}]{Note2}%
  \BibitemOpen
  \bibinfo {note} {We performed numerical simulations of the massive PVM using the velocity Verlet algorithm. In all the simulations, we numerically solved Eq.~\protect \eqref {eq:dimensionless EOM} with rescaled time $\protect \tilde {t}=t/\tau $ and length $\protect \tilde {\protect \bm {r}}_i=\protect \bm {r}_i/\sigma $, and set the time step $\Delta \protect \tilde {t}\leq 0.001$.}\BibitemShut {Stop}%
\bibitem [{Note3()}]{Note3}%
  \BibitemOpen
  \bibinfo {note} {By rescaling the function of $r$, such as Eq.~\protect \eqref {eq:analytical angle shift}, with $\tau _\xi $ and $\xi $ instead of $\tau $ and $\sigma $, new factors $\tau /\tau _\xi =\protect \frac {1}{2}(m_v/\pi \rho \xi ^2)$ and $\sigma /\xi =(m_v/\pi \rho \xi ^2)^{1/2}$ appear.Accordingly, this function has two variables, $r/\xi $ and $m_v/\pi \rho \xi ^2$. Note that $m_v/\pi \rho \xi ^2$ is an unknown quantity while $r/\xi $ is experimentally operable.}\BibitemShut {Stop}%
\bibitem [{\citenamefont {Takeuchi}\ \emph {et~al.}(2018)\citenamefont {Takeuchi}, \citenamefont {Kobayashi},\ and\ \citenamefont {Kasamatsu}}]{takeuchi2018doubly}%
  \BibitemOpen
  \bibfield  {author} {\bibinfo {author} {\bibfnamefont {H.}~\bibnamefont {Takeuchi}}, \bibinfo {author} {\bibfnamefont {M.}~\bibnamefont {Kobayashi}},\ and\ \bibinfo {author} {\bibfnamefont {K.}~\bibnamefont {Kasamatsu}},\ }\bibfield  {title} {\bibinfo {title} {{Is a doubly quantized vortex dynamically unstable in uniform superfluids?}},\ }\href {https://doi.org/10.7566/JPSJ.87.023601} {\bibfield  {journal} {\bibinfo  {journal} {J. Phys. Soc. Jpn.}\ }\textbf {\bibinfo {volume} {87}},\ \bibinfo {pages} {023601} (\bibinfo {year} {2018})}\BibitemShut {NoStop}%
\bibitem [{\citenamefont {Isoshima}(2008)}]{isoshima2008vortex}%
  \BibitemOpen
  \bibfield  {author} {\bibinfo {author} {\bibfnamefont {T.}~\bibnamefont {Isoshima}},\ }\bibfield  {title} {\bibinfo {title} {{Vortex chain structure in Bose-Einstein condensates}},\ }\href {https://doi.org/10.1143/JPSJ.77.094001} {\bibfield  {journal} {\bibinfo  {journal} {J. Phys. Soc. Jpn.}\ }\textbf {\bibinfo {volume} {77}},\ \bibinfo {pages} {094001} (\bibinfo {year} {2008})}\BibitemShut {NoStop}%
\bibitem [{\citenamefont {Aranson}\ and\ \citenamefont {Steinberg}(1996)}]{aranson1996stability}%
  \BibitemOpen
  \bibfield  {author} {\bibinfo {author} {\bibfnamefont {I.}~\bibnamefont {Aranson}}\ and\ \bibinfo {author} {\bibfnamefont {V.}~\bibnamefont {Steinberg}},\ }\bibfield  {title} {\bibinfo {title} {{Stability of multicharged vortices in a model of superflow}},\ }\href {https://doi.org/10.1103/PhysRevB.53.75} {\bibfield  {journal} {\bibinfo  {journal} {Phys. Rev. B}\ }\textbf {\bibinfo {volume} {53}},\ \bibinfo {pages} {75} (\bibinfo {year} {1996})}\BibitemShut {NoStop}%
\bibitem [{\citenamefont {Groszek}\ \emph {et~al.}(2018)\citenamefont {Groszek}, \citenamefont {Paganin}, \citenamefont {Helmerson},\ and\ \citenamefont {Simula}}]{groszek2018motion}%
  \BibitemOpen
  \bibfield  {author} {\bibinfo {author} {\bibfnamefont {A.~J.}\ \bibnamefont {Groszek}}, \bibinfo {author} {\bibfnamefont {D.~M.}\ \bibnamefont {Paganin}}, \bibinfo {author} {\bibfnamefont {K.}~\bibnamefont {Helmerson}},\ and\ \bibinfo {author} {\bibfnamefont {T.~P.}\ \bibnamefont {Simula}},\ }\bibfield  {title} {\bibinfo {title} {{Motion of vortices in inhomogeneous Bose-Einstein condensates}},\ }\href {https://doi.org/10.1103/PhysRevA.97.023617} {\bibfield  {journal} {\bibinfo  {journal} {Phys. Rev. A}\ }\textbf {\bibinfo {volume} {97}},\ \bibinfo {pages} {023617} (\bibinfo {year} {2018})}\BibitemShut {NoStop}%
\bibitem [{\citenamefont {Klimin}\ \emph {et~al.}(2015)\citenamefont {Klimin}, \citenamefont {Tempere}, \citenamefont {Lombardi},\ and\ \citenamefont {Devreese}}]{klimin2015finite}%
  \BibitemOpen
  \bibfield  {author} {\bibinfo {author} {\bibfnamefont {S.~N.}\ \bibnamefont {Klimin}}, \bibinfo {author} {\bibfnamefont {J.}~\bibnamefont {Tempere}}, \bibinfo {author} {\bibfnamefont {G.}~\bibnamefont {Lombardi}},\ and\ \bibinfo {author} {\bibfnamefont {J.~T.}\ \bibnamefont {Devreese}},\ }\bibfield  {title} {\bibinfo {title} {{Finite temperature effective field theory and two-band superfluidity in Fermi gases}},\ }\href {https://doi.org/10.1140/epjb/e2015-60213-4} {\bibfield  {journal} {\bibinfo  {journal} {Eur. Phys. J. B}\ }\textbf {\bibinfo {volume} {88}},\ \bibinfo {pages} {1} (\bibinfo {year} {2015})}\BibitemShut {NoStop}%
\bibitem [{\citenamefont {Klimin}\ \emph {et~al.}(2016)\citenamefont {Klimin}, \citenamefont {Tempere}, \citenamefont {Verhelst},\ and\ \citenamefont {Milo\ifmmode \check{s}\else \v{s}\fi{}evi\ifmmode~\acute{c}\else \'{c}\fi{}}}]{klimin2016finite}%
  \BibitemOpen
  \bibfield  {author} {\bibinfo {author} {\bibfnamefont {S.~N.}\ \bibnamefont {Klimin}}, \bibinfo {author} {\bibfnamefont {J.}~\bibnamefont {Tempere}}, \bibinfo {author} {\bibfnamefont {N.}~\bibnamefont {Verhelst}},\ and\ \bibinfo {author} {\bibfnamefont {M.~V.}\ \bibnamefont {Milo\ifmmode \check{s}\else \v{s}\fi{}evi\ifmmode~\acute{c}\else \'{c}\fi{}}},\ }\bibfield  {title} {\bibinfo {title} {{Finite-temperature vortices in a rotating Fermi gas}},\ }\href {https://doi.org/10.1103/PhysRevA.94.023620} {\bibfield  {journal} {\bibinfo  {journal} {Phys. Rev. A}\ }\textbf {\bibinfo {volume} {94}},\ \bibinfo {pages} {023620} (\bibinfo {year} {2016})}\BibitemShut {NoStop}%
\bibitem [{\citenamefont {Klimin}\ \emph {et~al.}(2018)\citenamefont {Klimin}, \citenamefont {Tempere},\ and\ \citenamefont {Milo{\v{s}}evi{\'c}}}]{klimin2018diversified}%
  \BibitemOpen
  \bibfield  {author} {\bibinfo {author} {\bibfnamefont {S.}~\bibnamefont {Klimin}}, \bibinfo {author} {\bibfnamefont {J.}~\bibnamefont {Tempere}},\ and\ \bibinfo {author} {\bibfnamefont {M.}~\bibnamefont {Milo{\v{s}}evi{\'c}}},\ }\bibfield  {title} {\bibinfo {title} {{Diversified vortex phase diagram for a rotating trapped two-band Fermi gas in the BCS-BEC crossover}},\ }\href {https://doi.org/10.1088/1367-2630/aaaceb} {\bibfield  {journal} {\bibinfo  {journal} {New J. Phys}\ }\textbf {\bibinfo {volume} {20}},\ \bibinfo {pages} {025010} (\bibinfo {year} {2018})}\BibitemShut {NoStop}%
\bibitem [{\citenamefont {Van~Alphen}\ \emph {et~al.}(2024)\citenamefont {Van~Alphen}, \citenamefont {Takeuchi},\ and\ \citenamefont {Tempere}}]{van2024splitting}%
  \BibitemOpen
  \bibfield  {author} {\bibinfo {author} {\bibfnamefont {W.}~\bibnamefont {Van~Alphen}}, \bibinfo {author} {\bibfnamefont {H.}~\bibnamefont {Takeuchi}},\ and\ \bibinfo {author} {\bibfnamefont {J.}~\bibnamefont {Tempere}},\ }\bibfield  {title} {\bibinfo {title} {{Splitting instability of a doubly quantized vortex in superfluid Fermi gases}},\ }\href {https://doi.org/10.1103/PhysRevA.109.043317} {\bibfield  {journal} {\bibinfo  {journal} {Phys. Rev. A}\ }\textbf {\bibinfo {volume} {109}},\ \bibinfo {pages} {043317} (\bibinfo {year} {2024})}\BibitemShut {NoStop}%
\bibitem [{\citenamefont {Patrick}\ \emph {et~al.}(2023)\citenamefont {Patrick}, \citenamefont {Gupta}, \citenamefont {Gregory},\ and\ \citenamefont {Barenghi}}]{Patrick2023stability}%
  \BibitemOpen
  \bibfield  {author} {\bibinfo {author} {\bibfnamefont {S.}~\bibnamefont {Patrick}}, \bibinfo {author} {\bibfnamefont {A.}~\bibnamefont {Gupta}}, \bibinfo {author} {\bibfnamefont {R.}~\bibnamefont {Gregory}},\ and\ \bibinfo {author} {\bibfnamefont {C.~F.}\ \bibnamefont {Barenghi}},\ }\bibfield  {title} {\bibinfo {title} {{Stability of quantized vortices in two-component condensates}},\ }\href {https://doi.org/10.1103/PhysRevResearch.5.033201} {\bibfield  {journal} {\bibinfo  {journal} {Phys. Rev. Res.}\ }\textbf {\bibinfo {volume} {5}},\ \bibinfo {pages} {033201} (\bibinfo {year} {2023})}\BibitemShut {NoStop}%
\bibitem [{\citenamefont {Hayashi}\ \emph {et~al.}(2013)\citenamefont {Hayashi}, \citenamefont {Tsubota},\ and\ \citenamefont {Takeuchi}}]{hayashi2013instability}%
  \BibitemOpen
  \bibfield  {author} {\bibinfo {author} {\bibfnamefont {S.}~\bibnamefont {Hayashi}}, \bibinfo {author} {\bibfnamefont {M.}~\bibnamefont {Tsubota}},\ and\ \bibinfo {author} {\bibfnamefont {H.}~\bibnamefont {Takeuchi}},\ }\bibfield  {title} {\bibinfo {title} {{Instability crossover of helical shear flow in segregated Bose-Einstein condensates}},\ }\href {https://doi.org/10.1103/PhysRevA.87.063628} {\bibfield  {journal} {\bibinfo  {journal} {Phys. Rev. A}\ }\textbf {\bibinfo {volume} {87}},\ \bibinfo {pages} {063628} (\bibinfo {year} {2013})}\BibitemShut {NoStop}%
\bibitem [{\citenamefont {Richaud}\ \emph {et~al.}(2023{\natexlab{b}})\citenamefont {Richaud}, \citenamefont {Lamporesi}, \citenamefont {Capone},\ and\ \citenamefont {Recati}}]{richaud2023mass-driven}%
  \BibitemOpen
  \bibfield  {author} {\bibinfo {author} {\bibfnamefont {A.}~\bibnamefont {Richaud}}, \bibinfo {author} {\bibfnamefont {G.}~\bibnamefont {Lamporesi}}, \bibinfo {author} {\bibfnamefont {M.}~\bibnamefont {Capone}},\ and\ \bibinfo {author} {\bibfnamefont {A.}~\bibnamefont {Recati}},\ }\bibfield  {title} {\bibinfo {title} {{Mass-driven vortex collisions in flat superfluids}},\ }\href {https://doi.org/10.1103/PhysRevA.107.053317} {\bibfield  {journal} {\bibinfo  {journal} {Phys. Rev. A}\ }\textbf {\bibinfo {volume} {107}},\ \bibinfo {pages} {053317} (\bibinfo {year} {2023}{\natexlab{b}})}\BibitemShut {NoStop}%
\bibitem [{\citenamefont {Huh}\ \emph {et~al.}(2024)\citenamefont {Huh}, \citenamefont {Mukherjee}, \citenamefont {Kwon}, \citenamefont {Seo}, \citenamefont {Hur}, \citenamefont {Mistakidis}, \citenamefont {Sadeghpour},\ and\ \citenamefont {Choi}}]{huh2024universality}%
  \BibitemOpen
  \bibfield  {author} {\bibinfo {author} {\bibfnamefont {S.}~\bibnamefont {Huh}}, \bibinfo {author} {\bibfnamefont {K.}~\bibnamefont {Mukherjee}}, \bibinfo {author} {\bibfnamefont {K.}~\bibnamefont {Kwon}}, \bibinfo {author} {\bibfnamefont {J.}~\bibnamefont {Seo}}, \bibinfo {author} {\bibfnamefont {J.}~\bibnamefont {Hur}}, \bibinfo {author} {\bibfnamefont {S.~I.}\ \bibnamefont {Mistakidis}}, \bibinfo {author} {\bibfnamefont {H.}~\bibnamefont {Sadeghpour}},\ and\ \bibinfo {author} {\bibfnamefont {J.-y.}\ \bibnamefont {Choi}},\ }\bibfield  {title} {\bibinfo {title} {{Universality class of a spinor Bose--Einstein condensate far from equilibrium}},\ }\href {https://doi.org/10.1038/s41567-023-02339-2} {\bibfield  {journal} {\bibinfo  {journal} {Nat. Phys.}\ }\textbf {\bibinfo {volume} {20}},\ \bibinfo {pages} {402} (\bibinfo {year} {2024})}\BibitemShut {NoStop}%
\bibitem [{\citenamefont {Huh}\ \emph {et~al.}()\citenamefont {Huh}, \citenamefont {Yun}, \citenamefont {Yun}, \citenamefont {Hwang}, \citenamefont {Kwon}, \citenamefont {Hur}, \citenamefont {Lee}, \citenamefont {Takeuchi}, \citenamefont {Kim},\ and\ \citenamefont {Choi}}]{huh2024beyond}%
  \BibitemOpen
  \bibfield  {author} {\bibinfo {author} {\bibfnamefont {S.}~\bibnamefont {Huh}}, \bibinfo {author} {\bibfnamefont {W.}~\bibnamefont {Yun}}, \bibinfo {author} {\bibfnamefont {G.}~\bibnamefont {Yun}}, \bibinfo {author} {\bibfnamefont {S.}~\bibnamefont {Hwang}}, \bibinfo {author} {\bibfnamefont {K.}~\bibnamefont {Kwon}}, \bibinfo {author} {\bibfnamefont {J.}~\bibnamefont {Hur}}, \bibinfo {author} {\bibfnamefont {S.}~\bibnamefont {Lee}}, \bibinfo {author} {\bibfnamefont {H.}~\bibnamefont {Takeuchi}}, \bibinfo {author} {\bibfnamefont {S.~K.}\ \bibnamefont {Kim}},\ and\ \bibinfo {author} {\bibfnamefont {J.-y.}\ \bibnamefont {Choi}},\ }\bibfield  {title} {\bibinfo {title} {{Beyond skyrmion spin texture from quantum Kelvin-Helmholtz instability}},\ }\href {https://doi.org/10.48550/arXiv.2408.11217} {\bibinfo  {journal} {arXiv:2408.11217}\ }\BibitemShut {NoStop}%
\bibitem [{\citenamefont {Lounasmaa}\ and\ \citenamefont {Thuneberg}(1999)}]{lounasmaa1999vortices}%
  \BibitemOpen
\bibfield  {journal} {  }\bibfield  {author} {\bibinfo {author} {\bibfnamefont {O.~V.}\ \bibnamefont {Lounasmaa}}\ and\ \bibinfo {author} {\bibfnamefont {E.}~\bibnamefont {Thuneberg}},\ }\bibfield  {title} {\bibinfo {title} {{Vortices in rotating superfluid 3He}},\ }\href {https://doi.org/10.1073/pnas.96.14.7760} {\bibfield  {journal} {\bibinfo  {journal} {Proc. Natl. Acad. Sci. USA}\ }\textbf {\bibinfo {volume} {96}},\ \bibinfo {pages} {7760} (\bibinfo {year} {1999})}\BibitemShut {NoStop}%
\bibitem [{\citenamefont {Jackson}(2021)}]{jackson2021classical}%
  \BibitemOpen
  \bibfield  {author} {\bibinfo {author} {\bibfnamefont {J.~D.}\ \bibnamefont {Jackson}},\ }\href@noop {} {\emph {\bibinfo {title} {{Classical Electrodynamics}}}}\ (\bibinfo  {publisher} {John Wiley \& Sons, New York},\ \bibinfo {year} {2021})\BibitemShut {NoStop}%
\end{thebibliography}%

\end{document}